\documentclass[a4paper]{aa} % for papers
 \usepackage{amsmath,ifpdf,txfonts,natbib,graphicx}
 \usepackage{array}
 \usepackage[english]{babel}
 \bibpunct{(}{)}{;}{a}{}{,} % to follow the A&A style

 \ifpdf
   \usepackage[pdftex,unicode,pdfauthor=Yvart,colorlinks]{hyperref}
 \else
   \usepackage[dvips,unicode,pdfauthor=Yvart,colorlinks]{hyperref}
 \fi
\newcommand{\eab}{\begin{array}} \newcommand{\eae}{\end{array}}
\newcommand{\eqab}{\begin{eqnarray}} \newcommand{\eqae}{\end{eqnarray}}
\newcommand{\eqb}{\begin{equation}} \newcommand{\eqe}{\end{equation}}

\newcommand{\sub}{\mathrm{sub}}

\newcommand{\water}{\mathrm{H}_\mathrm{2}\mathrm{O}}

\newcommand{\HH}{\mathrm{H}_\mathrm{2}}

\newcommand{\Macc}{$\dot M_{\rm acc}$}
\newcommand{\Msun}{$M_\odot$}
\newcommand{\Mstar}{$M_\star$}
\newcommand{\romax}{$r_{0}^\mathrm{max}$}

\newcommand{\Lbol}{$L_{\rm bol}$}
\newcommand{\Lacc}{$L_{\rm acc}$}
\newcommand{\Lstar}{$L_\star$}
\newcommand{\Tkin}{$T_\mathrm{kin}$}
\newcommand{\Tdust}{$T_\mathrm{d}$}
\newcommand{\Tbol}{$T_\mathrm{bol}$}
\newcommand{\Rsub}{$R_\sub$}

\newcommand{\TdV}{$\int TdV$\,}

\newcommand{\mMstar}{M_\star}
\newcommand{\mMacc}{\dot M_{\rm acc}}
\newcommand{\mMsun}{M_\odot}
\newcommand{\Rsun}{R_\odot}
\newcommand{\Lsun}{L_\odot}

\newcommand{\Rstar}{R_\star}
\newcommand{\Tstar}{T_\star}

\newcommand{\UMacc}{\mathrm{M_\odot}~\mathrm{yr}^{-1}}

\newcommand{\nH}{$n_\mathrm{H}$}

\newcommand{\nff}{$n^\mathrm{ff}_{\mathrm H_2}$}
\newcommand{\nenv}{$n^\mathrm{env}_{\mathrm H_2}$}

\newcommand{\AU}{\mathrm{AU}}

\newcommand{\cm}{$\mathrm{cm}^{-3}$}
\newcommand{\dr}{\mathrm{d}}

\newcommand{\erg}{\mathrm{erg}}
\newcommand{\eV}{\mathrm{eV}}

\newcommand{\GHz}{\mathrm{GHz}}

\newcommand{\K}{\mathrm{K}}

\newcommand{\N}{\mathrm{N}}

\newcommand{\pc}{\mathrm{pc}}

\newcommand{\s}{\mathrm{s}}

\newcommand{\yr}{\mathrm{yr}}

\newcommand{\kms}{\rm{km~s}$^{-1}$} 
 
 \titlerunning{Water line profiles in MHD disk winds}
 \authorrunning{Yvart et al.}       %\shortauthors{Panoglou et al.}
\begin{document}

% --- Title ---
\title{Molecule survival in magnetized protostellar disk winds} 
\subtitle{II. Predicted H$_2$O line profiles versus {\em Herschel}/HIFI observations}
%\subtitle{}

% --- Authors ---
\author{W.~Yvart\inst{1,2,3}
   \and  S.~Cabrit\inst{1,3,4}
   \and  G.~Pineau des For\^ets\inst{5,1}
%    \and  P.~J.~V.~Garcia\inst{2,4,5}
    \and  J.~Ferreira\inst{4}
%    \and  F.~Casse\inst{6}
}

\offprints{\tt walter.yvart@obspm.fr}

 \institute{LERMA, Observatoire de Paris, PSL Research University, CNRS, UMR8112, F-75014 Paris, France
   \and    LESIA, Observatoire de Paris, PSL Research University, CNRS, F-92190 Meudon, France
   \and Sorbonne Universit\'es, UPMC Univ. Paris 06, F-75005, Paris, France
   \and     IPAG, UMR 5521 du CNRS, Observatoire de Grenoble, 
            38041 Grenoble Cedex, France
   \and    IAS, UMR 8617 du CNRS, Universit\'e de Paris-Sud,
             91405 Orsay, France
%   \and     Laboratoire Astroparticule \& Cosmologie, Universit\'e
%             Paris 7, UMR 7164 du CNRS, 75205 Paris Cedex 13, France
}

% --- Date ---
   \date{}

% --- Abstract ---
\abstract %5{}needed
%context
{The origin of molecular protostellar jets and their role in extracting angular
momentum from the accreting system are important open questions
in star formation research. In the first paper of this series 
we showed that a dusty magneto-hydrodynamic (MHD) disk wind 
appeared promising to explain the pattern of H$_2$ temperature and collimation in the youngest jets.}
%Aims
{We wish to see whether the high-quality H$_2$O emission profiles of low-mass protostars,
        observed for the first time by the HIFI spectrograph 
        on board the Herschel satellite, remain consistent with the MHD disk wind hypothesis, and which constraints they would set on the underlying disk properties.}
%Methods
{We present synthetic H$_2$O line profiles predictions for a typical MHD disk wind solution with various values of disk accretion rate, stellar mass, extension of the launching area, and view angle. We compare them in terms of  line shapes and intensities with the HIFI profiles observed by the WISH Key Program towards a sample of 29 low-mass Class 0 and Class 1 protostars.}
%Results
{A dusty MHD disk wind launched from 0.2--0.6 AU AU to 3--25 AU can reproduce to a remarkable degree the observed shapes and intensities of 
        the broad H$_2$O component observed in low-mass protostars, both in the fundamental 557 GHz line and in more excited lines. Such a model also readily reproduces the observed correlation of 557 GHz line luminosity with envelope density, if the infall rate at 1000 AU is 1--3 times the disk accretion rate in the wind ejection region. It is also compatible with the typical disk size and bolometric luminosity in the observed targets. However, the narrower line profiles in Class 1 sources suggest that MHD disk winds in these sources, if present, would have to be slower and/or less water rich than in Class 0 sources. }
%Conclusions
{MHD disk winds appear as a valid (though not unique) option to consider for the origin of the broad H$_2$O component in low-mass protostars. ALMA appears ideally suited to further test this model by searching for resolved signatures of the warm and slow wide-angle molecular wind that would be predicted.}

%TODO figure TdV-nH et TdV-Lbol
%TODO check if class 1 = Macc and class0 = 3Macc
%TODO: write abstract
%TODO add comment on rout truncation
%TODO acknowledgements
%TODO:add refs disks, M*, Lbol, i

% --- Keywords ---

\maketitle

%--------------------------------------------------------------------------------------------------%
\section{Introduction}\label{section_intro}
%--------------------------------------------------------------------------------------------------%

Supersonic bipolar jets are ubiquitous in young accreting stars from the initial deeply embedded phase (called Class 0) -- where most of the mass still lies in the infalling envelope -- through the late infall phase (Class 1) -- where most of the mass is in the central protostar -- to optically revealed young stars (Class 2 or T Tauri phase) -- where the envelope has dissipated, but residual accretion continues through the circumstellar disk. The similarities in jet collimation, variability timescales, and ejection efficiency among these three phases suggests that robust universal processes are at play. In particular, the high kinetic power estimated in these jets (at least 10\% of the accretion luminosity in low-mass sources) and the evidence that they are collimated on very small scales $\le$ 20--50 AU suggests that {  magneto-hydrodynamic (MHD)} processes play a key role and that jets could remove a significant fraction of the gravitational energy and excess angular momentum from the accreting system \citep[see e.g.][for a recent review]{Frank2014}. However, it is still unclear which parts of the MHD jet arise from a stellar wind \citep{Sauty94,Matt08}, the interaction zone between stellar magnetosphere and inner disk edge \citep{Shu94,Zanni2013} {  and/or} the Keplerian disk surface \citep[hereafter D-wind,][]{BP82,KP00}. Global semi-analytical solutions, confirmed by numerical simulations, have shown that the presence and radial extent of steady magneto-centrifugal D-winds in young stars depend chiefly on the distribution of vertical magnetic fields and diffusivity in the disk \citep[see e.g.][]{CasseFerreira2000,SFendt2012}. Both of these quantities have a fundamental impact on the disk angular momentum extraction, turbulence level, and planet formation and migration \citep[see reviews by][]{Turner2014,Baruteau2014}. {  Finding observational diagnostics that can establish or exclude the presence of MHD D-winds in young stars is thus essential not only to pinpoint the exact origin of protostellar jets, but also to better understand disk accretion and planet formation.} 

%and even represent the first clear manifestation of a newly born protostar. The high kinetic power estimated in these jets (at least 10\% of the accretion luminosity in low mass sources) and the evidence that they are magnetically collimated on small scales (20--50 AU; Ray et al. 2007; Cabrit et al. 2007) suggests that they could remove a significant fraction of the gravitational energy and excess angular momentum from the accreting system. However it is yet unclear {\it which parts of the jet} are ejected from the stellar photosphere (Matt et al.), the stellar magnetosphere-disk interaction zone (Zanni \& Ferreira 2013), and the surface of the protoplanetary disk (Disk Wind; Blandford \& Payne, Ferreira). Global semi-analytical solutions, confirmed by numerical simulations, have shown that the presence and radial extent of DWs in young stars depend chiefly on the distribution of the vertical magnetic field and diffusivity in the disk (REFs). Both of these quantities have fundamental impact on the process of angular momentum extraction from the disk, its level of  turbulence, and the conditions of planet formation and migration (REFs). Finding reliable diagnostics for the presence and radial extent of DWs in young stars is thus important not only to pin-point the origin of protostellar jets but also to better understand the physics and evolution of protoplanetary disks. 

{  Possible signatures of jet rotation consistent with steady D-winds launched from 0.1-30 AU have been claimed in several atomic and molecular protostellar jets, using HST or millimeter interferometers \citep[see e.g.][for a recent review]{Frank2014}. However, the resolution currently achieved across the jet beam is not yet sufficient to fully disentangle other effects such as asymmetric shocks or jet precession / orbital motion. Another independent and} powerful {  diagnostic of the possible}  presence and radial extent of an extended D-wind is the molecular richness and temperature of the flow. In the first paper of this series \citep{Panoglou2012}, we investigated the predicted thermo-chemical structure of a steady self-similar dusty D-wind around low-mass protostars of varying accretion rates, taking into account the wind irradiation by far-ultraviolet (FUV) and X-ray photons from the accretion shock and the stellar corona. We confirmed the pioneering insight of \citet{Safier1993a} that H$_2$ molecules can survive beyond some critical launch radius $r_c \simeq 0.2-1$ AU, and showed that the wind is heated up to $T \simeq 500-2000$ K by ambipolar ion-neutral friction with both $r_c$ and $T$ increasing with decreasing wind density. The predicted evolutions of molecular jet temperature, collimation, and speed with evolutionary stage were found qualitatively consistent with existing observational trends. We also predicted a high abundance of H$_2$O along dusty streamlines launched from 1 AU. This emerged as a natural consequence of the D-wind heating by ion-neutral drag, which allowed quick gas-phase reformation of H$_2$O by endothermic reactions of O and OH with H$_2$ to balance FUV photodissociation. {   This temperature-dependence of H$_2$O abundance, combined with a more density-sensitive line excitation than for CO (due to a much larger dipole moment) means that H$_2$O emission appears as a particularly promising discriminant diagnostic of extended dusty D-winds in young low-mass stars.}

{  Thanks to the exquisite sensitivity of the HIFI spectrometer on board the {\it Herschel} satellite, high-quality line profiles of H$_2$O emission 
were obtained for the first time towards a wide sample of low-mass protostars and their outflows. Off-source spectra are dominated by localized shocked regions created by time variable ejection and interaction with the ambient medium \citep[see e.g.][]{Lefloch2012,Santangelo2012}. Hence their varied line shapes trace disturbed flow dynamics that do not reflect the initial wind structure. In contrast, on-source spectra encompass only the innermost $\simeq 5000$ AU (at the distance of Perseus), which may still carry the imprint of the initial wind structure established by the acceleration process. On-source H$_2$O HIFI spectra were acquired towards 29 low-mass protostars,
in the course of the WISH Key Program \citep{WISH}. They differ from off-source spectra in revealing the presence of three roughly Gaussian velocity components at the flow base, with common properties across the whole sample
%\footnote{The dynamics and line profiles in the innermost 5000 AU differ from the "pure outflow" positions well offset from the source, which are dominated by localized shock structures in the outflow and tend to present exponential --- rather than gaussian --- line wings \citep[see e.g.][]{Lefloch2012}. We concentrate here on HIFI spectra obtained close to the source, where a steady disk-wind, if present, could dominate the emission.} 
\citep{Kristensen2012557GHz,Mottram2014}: a narrow central component arising from the infalling envelope; a medium component,  slightly offset to the blue and detected only in a few sources, interpreted as a dissociative compact ``spot-shock'' between a wide-angle wind and the swept-up outflow cavity  \citep{Jshocks-Kristensen}; and a ubiquitous broad component (BC) centered near systemic velocity and extending up to $\pm 50$ \kms, {  which usually dominates the total H$_2$O flux and is still of  unclear origin. Low-$J$ CO line profiles in the same sources only start tracing the BC for $J =10$ owing to a much stronger contribution in CO from the envelope and the cool swept-up outflow cavity  \citep{Yildiz2013}. Hence, H$_2$O profiles appear as the current best specific tracer for constraining the physical origin of the BC, and we concentrate on this tracer in the following. Average excitation conditions and emission sizes for the BC were recently estimated by \citep{Mottram2014} based on a multi-line H$_2$O analysis, and an origin in thin irradiated C-shocks along the outflow cavity walls was proposed. A full chemical-dynamical model of this situation with line profile predictions remains to be developed to test this hypothesis. An alternative worth considering is that the BC in H$_2$O might trace dusty MHD D-winds heated by ambipolar diffusion (the same heating process as in magnetized C-shocks), since those winds are also naturally expected to be water-rich for the reasons noted above, and to cover a broad range of flow velocities (due to their broad range of launch radii). Testing this alternative scenario against BC observations is an important goal, as it may set strong constraints on ill-known fundamental properties of  Keplerian disks in protostars such as their size, accretion rate, and magnetization, which are central elements  for modern theories of disk formation and magnetic braking in star formation \citep[see][for a recent review]{Li-PPVI}.}
%Since H$_2$O is more difficult to synthesize and excite in the wind than CO, we concentrate on this tracer in the following as a first step. 

In the present paper, we thus go one step further in exploring observational diagnostics of steady, self-similar dusty MHD D-winds by presenting synthetic H$_2$O line profiles and comparing them to HIFI /{\em Herschel} observations of young low-mass Class 0 and Class I sources.  {  Our specific goals are to see (1) whether  a typical MHD D-wind is able to reproduce the range of H$_2$O line profile shapes and intensities of the BC  --- an excessive predicted flux could argue against the presence of extended molecular D-winds in these sources --- and (2) which range of launch radii and disk accretion rates would be implied if a D-wind were indeed the dominant contributor to the H$_2$O BC.} Section 2 describes the model improvements made since Paper~I and the parameter space explored. Section 3 presents the calculated temperature and water gas-phase abundance in our D-wind model, and the effect of various system parameters on the emergent H$_2$O line profiles. Section 4 {  demonstrates the excellent agreement between} our synthetic line predictions and HIFI {  on-source spectra} of low-mass protostars in terms of line profile {  shapes, absolute and relative intensities, and correlation with} envelope density. Section 5  {  verifies whether the D-wind model explored here for the origin of the BC would also be} consistent with independent observational constraints on disk size, {  accretion luminosity}, and inclination in low-mass protostars. Section 6 summarizes our main conclusions and future prospects. 

%Young embedded protostars are seen to drive powerful collimated winds (jets), which are invoked to solve
%several key puzzles in star formation: the removal of excess angular momentum from accreting matter (by the so-called
%magneto-centrifugal mechanism),  the limited core to star efficiency of 30\% (by re-ejecting a fraction of infalling gas
%or dissipating the envelope), the low overall efficiency of star formation in molecular clouds (through turbulence injection by the outflows). 
%Confirming these hypotheses requires to pin-point the launching mechanism of these outflows, which determines how much mass and angular momentum they extract; and their latitude distribution in mass, momentum, and kinetic energy flux, which determines how they interact with surrounding matter.
%One hypothesis is that a large fraction of the matter is ejected from the surface of magnetized accretion disks.

%--------------------------------------------------------------------------------------------------%
\section{Model description}\label{section_model}
%--------------------------------------------------------------------------------------------------%

%------------------------------------------------------%
\subsection{Dynamics and thermo-chemistry} \label{model_mdh}

Our modeling work is based on the thermo-chemical model of dusty MHD disk winds developed in \citep[][hereafter Paper I]{Panoglou2012}, where all the model
ingredients are presented in detail. We only give a brief summary of the main elements here, and present the improvements made in order to obtain line profile predictions in H$_2$O comparable to {\it Herschel}/HIFI observations. 

The wind dynamics are prescribed using a self-similar, steady, axisymmetric MHD solution of accreting-ejecting Keplerian disks  computed by \citet{CasseFerreira2000} with a prescribed heating function at the disk surface (known as ``warm" disk wind solution). It assumes a thermal pressure scale-height $h = 0.03r$ (where $r$ denotes the cylindrical radius), and has a magnetic lever arm parameter $\lambda \simeq (r_A/r_0)^2 = 13.8$,  where $r_A$ is the (cylindrical) radius of the wind Alfv\'en point along the magnetic surface anchored at $r_0$ in the disk midplane. This particular solution was chosen because (i) the small $\lambda$ value gives a good match to the tentative rotation signatures reported at the base of the DG Tauri atomic jet \citep{Pesenti2004} and of other T Tauri jets \citep{Ferreira2006} and (ii) the higher wind mass-loading ($\xi \simeq  1/(2\lambda-2) \simeq 0.04$) is more consistent with observed atomic jet mass-fluxes than vertically isothermal ``cold'' solutions with large $\lambda$ values \citep{Garcia2001b,Ferreira2006}. The chosen solution provides the non-dimensional streamline shape and distribution of velocity, density, and Lorentz force along it as a function of polar angle $\theta$. The 2D density distribution and streamlines for typical source parameters are illustrated in Fig.~1 of Paper~I. 

Because the dynamical timescales for crossing the inner few 1000 AU of the jet are short,
\begin{math}
t_{\mathrm{dyn}} \sim 100 \, \yr \times ({r_0}/{\AU})^{1.5},
\end{math}
the gas temperature, ionization level, and chemical abundances in the disk wind are generally out of equilibrium. 
The time-dependent thermo-chemistry of the gas  must therefore be integrated numerically along each streamline as a function of altitude $z$ above the midplane.
The code performing this calculation is based on the C-shock model of \citet{Flower2003} with additional treatment for the effect of stellar coronal X-rays and of FUV photons from the accretion hot spots (attenuated by gas and dust on the inner streamlines of the disk wind). The various heating, cooling, and chemical processes along the dusty streamlines are described and discussed in detail in Paper~I. 

In order to obtain reliable predictions in H$_2$O lines, the thermo-chemical model from Paper~I was upgraded in several ways:
\begin{enumerate}
\item Self-consistent, non-local calculation of $\HH$ and CO self-shielding 
          (see Appendix~\ref{app:shielding}).
\item Time-dependent integration of NLTE rotational level populations of $\water$ and the corresponding gas heating/cooling (see Appendix~\ref{app:lines}).   
\item Calculation of dust temperature including the dust opacity to its own radiation (see Appendix~\ref{app:model_DUSTY}).
\item Refined initial chemical abundances at the base of streamlines (see Appendix~\ref{app:initial_abun}).    
\end{enumerate}

As in Paper~I, we will assume that the {atomic} MHD disk wind starts at a typical inner disk truncation radius of 0.07~AU (where it starts to provide gas extinction against coronal X-rays), while the first calculated {molecular} streamline is launched from the dust sublimation radius, \Rsub. The thermo-chemical evolution is then calculated for a series of streamlines with launch radii $r_0$ ranging from \Rsub\ to a freely specified maximum value \romax $\le 25 \AU$. 
%We have verified that ion-neutral drift remains sufficiently small for the single-fluid approximation to be valid in this region. 
%We assume that the wind dynamics inside \romax\ are not severely affected by this outer truncation \citep[as concluded by][from their MHD simulations]{Stute}. 
The thermo-chemical integration starts just above  the disk surface at the wind slow point, located at $z \simeq 1.7h \simeq 0.05 r_0$, and stops at $z \simeq 1000 r_0$ ($z/r \simeq 20$) where streamlines start to bend inwards. A recollimation shock may form beyond this point, and the steady dynamical solution would no longer be valid. 

The parameters describing the stellar radiation field remain as in Paper I, where the reader is referred  for a justification of these assumptions:  (1) a stellar photosphere of effective temperature $\Tstar = 4000 \, \K$ and radius $\Rstar = 3 \Rsun$, producing a luminosity \Lstar $= 2.1 \Lsun$; (2) a hard thermal X-ray spectrum of characteristic energy $k_B T_X = 4 \, \K \eV$ and luminosity $L_\mathrm{X} = 10^{30} \, \erg \s^{-1}$, typical of active coronae in young low-mass stars; and (3) an accretion hot spot emitting as a blackbody of temperature $T_\mathrm{hs}$ = 10\,000~K and luminosity $L_{\mathrm{hs}}$ equal to half of the accretion luminosity $L_\mathrm{acc} = G$ \Mstar\Macc $/R_\star$. The hot-spot luminosity is thus the only radiative contribution that will vary from model to model (see Table~\ref{tab:models}).

%------------------------------------------------------%
\subsection{Model grid} \label{sec:model_grid}

\noindent

\begin{table}
\caption{Models in order of increasing source accretion age and decreasing wind density.}
\label{tab:models}
\centering
\begin{tabular}{r r r r r r }
\hline\hline
% 1         2                       3                                        4                                           5               6  
\Mstar\ & \Macc\  & $t_{\rm acc}$ & \nff(1000 AU)  & $L_{\mathrm{hs}}$ &  \Rsub   \\%  & $L_{\mathrm{tot}}$\\
(\Msun) & ($\UMacc$) & yr & (cm$^{-3}$) & ($\Lsun$) & ($\AU$)  \\
\hline
\multicolumn{4}{l}{\smallskip model X0 = Extreme Class 0} \\
0.1 & $2 \times 10^{-5}$  & $5 \times 10^3$ & $2.3 \times 10^6$  &  10.5 & 0.63     \\%     &  23   \\
\hline
\multicolumn{4}{l}{\smallskip  \bf model S0 = Standard Class 0 (reference model)} \\
\smallskip
\textbf{0.1} & $\mathbf{5 \times 10^{-6}}$ & $\mathbf{2 \times 10^4}$  & $\mathbf{6 \times 10^5}$  &  {\bf 2.6} &  {\bf 0.31}   \\%    &  7.2   \\
\hline
\multicolumn{4}{l}{\smallskip model X1 = Extreme Class 1} \\
0.5 &  $5 \times 10^{-6}$   &  $10^5$  & $2.5 \times 10^5$    &  13.1  & 0.64       \\
\hline
\multicolumn{4}{l}{\smallskip model S1= Standard Class 1} \\
0.5 & $10^{-6}$    &  $5 \times 10^5$   & $5 \times 10^4$    &  2.6  & 0.27          \\%     &  23   \\
\hline
\end{tabular}
\tablefoot{
%\Mstar\ is the stellar mass, \Macc\ is the disk accretion rate,
%$\Rsub = r_0^\mathrm{min}$ is the dust sublimation radius, and
%\romax\ the maximal launch radius,
%and $i$ the inclination angle to line of sight. 
\Mstar\ is the stellar mass, \Macc\ the accretion rate through the disk, and $t_{\rm acc} \equiv$ \Mstar/\Macc\ the characteristic accretion age. 
\nff(1000 AU) is the H$_2$ density at 1000 AU of an envelope in free fall at rate \Macc\ onto a star of mass \Mstar. It scales as \Macc\Mstar$^{-0.5}$, like the wind density field, and thus acts as a proxy for the latter quantity. $L_{\mathrm{hs}} \equiv 0.5 G$\Mstar\Macc/$\Rstar$ is the accretion hot-spot luminosity and
\Rsub\ the dust sublimation radius as calculated by DUSTY (see Appendix~\ref{app:model_DUSTY}). For each source model, synthetic line profiles were calculated for a grid of maximum launch radii \romax = 3.2, 6.4, 12.8, and 25.6 au and inclination angles $i =$ 30\degr, 60\degr, and 80\degr\ from pole-on. }
%Other constant parameters are~: $\Tstar = 4000 \, \K$, hot spot temperature $T_\mathrm{hs}$ = $10\,000$ $\K$,
%The stellar luminosity $\Lstar$ is fixed to 2.1 $\Lsun$ (see text for other fixed parameters).}
\end{table}

The 2D spatial distribution of density, velocity and electric current values in the disk wind is obtained from the non-dimensional MHD solution by prescribing two free dimensional parameters: the stellar mass, \Mstar, and the accretion rate in the disk ejecting zone, \Macc. Each pair of values (\Mstar,\Macc) defines a so-called ``source model". Because the stellar mass and disk accretion rate are still poorly known in most protostars, and each combination requires a large number of CPU-consuming thermo-chemical calculations, we only explored a restricted grid of four source models summarized in Table~\ref{tab:models}, meant to represent typical parameters for low-luminosity Class 0 and Class 1 sources:
\begin{itemize}
\item two values of stellar mass:  Following Paper~I, we adopt \Mstar = 0.1 \Msun\ for our Class 0 models, and 
 \Mstar = 0.5 \Msun\ for our Class 1 models (where the star has accreted most of its final mass). 
%(described in section~\ref{results_class}),
\item two values of accretion rate for each stellar mass: For the Class 0 models we consider $5 \times 10^{-6} \UMacc$
(standard Class 0  from Paper I = S0 model) and $2 \times 10^{-5} \UMacc$ (extreme Class 0 = X0 model). For the Class 1 sources we consider $10^{-6} \UMacc$ (standard Class 1 from Paper I = S1 model), and $5 \times 10^{-6} \UMacc$ (extreme Class 1 = X1 model). 
%described in section~\ref{results_macc}.
\end{itemize}

The range of  \Macc\ values in our Class 0 and Class 1 models is in line with the predictions of recent 3D resistive MHD simulations of collapsing low-mass protostars \citet[see Fig. 7 in][]{Machida2013}. 
The corresponding characteristic accretion ages $t_{\rm acc}$ = \Mstar/\Macc\ are $0.5-2\times10^4$ yr for the Class 0 models and $1-5\times10^5$ yr for the Class 1 models. 

We note that because the model is self-similar, the wind velocity field scales as ${M_\star}^{0.5}$ (Keplerian law), while the wind density field scales as \nH $\propto$ \Macc $M_\star^{-0.5}$. The latter scaling is the same as for an envelope in free fall at rate \Macc\ onto a point mass \Mstar. Therefore, we can characterize how dense each wind model is by using  the density such a fiducial envelope would have at  1000 AU as a proxy. This parameter, which we denote as \nff(1000 AU), is listed in Col. 4 of Table~\ref{tab:models} and shows that
the wind density drops by a factor $\simeq$ 45 in total from model X0 to model S1. Table~\ref{tab:models} also lists the  hot-spot luminosity $L_{\mathrm{hs}} = 0.5L_{\mathrm{acc}}$ for each source model, and the corresponding dust sublimation radius \Rsub\ calculated  as described in Appendix~\ref{app:model_DUSTY}. 
%{  SC: add something here on the few values of \Mstar\ determined from keplerian disk rotation}

For each source model, the synthetic emergent line profiles depend on two additional free parameters, namely the maximum radial extension \romax\ of the emitting MHD disk wind, and the inclination angle $i$ of the disk axis to the line of sight. For the present exploratory study, we restricted ourselves to the following coarse grid : 
\begin{itemize}
\item  four values of \romax\ ~: 3.2, 6.4, 12.8 and 25 AU. %described in section~\ref{results_r0max},
\item three values of the inclination angle $i$~: near pole-on (30\degr), median (60\degr) and near edge-on (80\degr). %described in section~\ref{results_angle}.
\end{itemize}

%\NEW{Two additional class 1 models (described in section~\ref{results_class}),
%where the star has accreted most of its final mass,
%have been further defined. A first with an
%accretion rate $\Macc = 10^{-6} \, \UMacc$, which corresponds
%to standard "class 1" model of Paper I (i. e. $\Mstar = 0.5 \, \Msun$),
%and a model with $\Macc = 5 \times 10^{-6} \, \UMacc$.
%This higher accretion rate correspond to the reference class 0 model
%evolved into class 1 with constant accretion rate,
%allowing thereby to prospect influence of the age of the protostar.}

%Other constant parameters are summarized in table~\ref{table_model} caption.

%The class 0 higher accretion rate model ($\Macc=2 \times 10^{-5} \, \UMacc$)
%has densities 4 times greater than our reference model, and a
%twice larger dust sublimation radius $\Rsub = 0.63 \, \AU$.

%The reference model is
%constituted of a series of streamlines launched from~: $\Rsub = 0.31 \, \AU$,
%0.4, 0.6, 0.8, 1, 1.2, 1.6, 3.2, 6.4, 9.6, 12.8, 16, 19.2, 22.4 and 25~$\AU$.

%\subsection{Calculation of synthetic H$_2$O line profiles}
By varying the four free parameters (\Mstar,\Macc,\romax, $i$) we thus obtain a grid of $ 2 \times 2 \times 4 \times 3 = 48$ synthetic line profiles for each H$_2$O line and source distance $d$. 
Synthetic channel maps are computed using the same method as in \citet{Cabrit1999,Garcia2001b} for optical forbidden lines, except that the line emissivity at each $(r,z,\phi)$ is now multiplied by the escape probability {in the observer's direction} to account for the large H$_2$O line opacity and its anisotropy (see Appendix~\ref{app:lines} ). Channel maps are then %converted to brightness temperature scale in K, and 
convolved by the {\em Herschel} beam at the line frequency for the specified source distance, and the flux towards the central source in each channel map is used to build the predicted line profile as a function of radial velocity. 
%is obtained by plotting the beam-averaged brightness temperature at the central source position as a function of channel map velocity. We thus obtain a grid of $ 2 \times 2 \times 4 \times 3 = 48$ synthetic line profiles for each H$_2$O line and source distance $d$. % (see appendix~\ref{profiles}). 

%--------------------------------------------------------------------------------------------------%
\section{Model results}\label{section_results}
%--------------------------------------------------------------------------------------------------%

%---------------------------------------------------------%
%\subsection{Dust temperature} \label{results_Tdust}

%The Figure~\ref{FIGURE1}
%shows gas temperature \Tkin\
%and $\water$ fractional abundance to \nH\
%in meridional cross-section of disk wind.
%The lower panel zooms into the innermost part of the disk wind.

\subsection{Gas temperature} \label{results_Tgas}

%TEMPERATURE

%Ion-neutral drag is the main heating term except at the wind base, where heat transfer from dust and collisional de-excitation of pumped H$_2$O levels dominate.

Figure~\ref{fig:temp_cuts} presents color maps of the calculated distribution of gas temperature in our four wind models. Two-dimensional interpolation was applied between the discrete calculated streamlines to fill the $r,z$ plane. As shown by \citet{Safier1993a,Garcia2001a} and Paper~I, the main heating term is ion-neutral drag (denoted $\Gamma_{\rm drag}$ hereafter). This process arises because the neutral fluid cannot feel  the Lorentz force that accelerates the MHD disk-wind directly, and thus lags behind the accelerated charged particles. This velocity drift generates ion-neutral collisions that drag along and accelerate the neutrals, but also degrade the kinetic energy of drift motions into heat. It is the very same process that dissipates kinetic energy and provides gas heating in C-type shocks. Hence there are definite analogies between the thermal (and chemical) structure of molecular MHD disk-winds and that of C-shocks, as noted in Paper ~I, except that the heating rate and spatial scale are set here by the dynamics of MHD wind acceleration instead of the velocity jump in C-shocks. 

It can be seen in Figure~\ref{fig:temp_cuts} that overall the wind  gets warmer from right to left in this figure as \Macc\ drops and \Mstar\ increases. This stems from the way $\Gamma_{\rm drag}$ scales with \Macc\ and \Mstar\ and adjusts to balance adiabatic and radiative cooling (see Paper~I). Typical temperatures on intermediate streamlines are $\simeq 100$~K in the extreme Class 0,  $\simeq 300$~K in the standard Class 0, $\simeq 600$~K in the extreme Class 1, and $\simeq 1000$~K in the standard Class 1. Within a given source model, the gas temperature decreases while $r_0$ increases, again because of the scaling of $\Gamma_{\rm drag}$ with $r_0$ (see Paper~I). The slightly higher dust temperature compared with Paper I, due to the inclusion of photon-trapping, only affects the gas temperature at the dense wind base where gas and dust are well coupled. 
%Compared to Paper~I, the higher dust temperature resulting from dust opacity to its own radiation (see Appendix~\ref{app:model_DUSTY}) makes the gas warmer near the wind sonic point, where gas-dust coupling is efficient. Radiative pumping by the dust diffuse background further increases the gas temperature at the wind base through collisional de-excitation of pumped H$_2$O levels, although the effect remains moderate  (see Appendix~\ref{app:lines}). %In the S0 streamline shown in  Figure~\ref{fig:temp_graph}, it varies only by a factor $\leq$ 3 from start to end (with a maximum at the sonic point and at $z/r \simeq 60$, and a minimum at $z/r \simeq 3$ and at the recollimation point). 

\begin{figure}
 \includegraphics[angle=0, width=\linewidth]{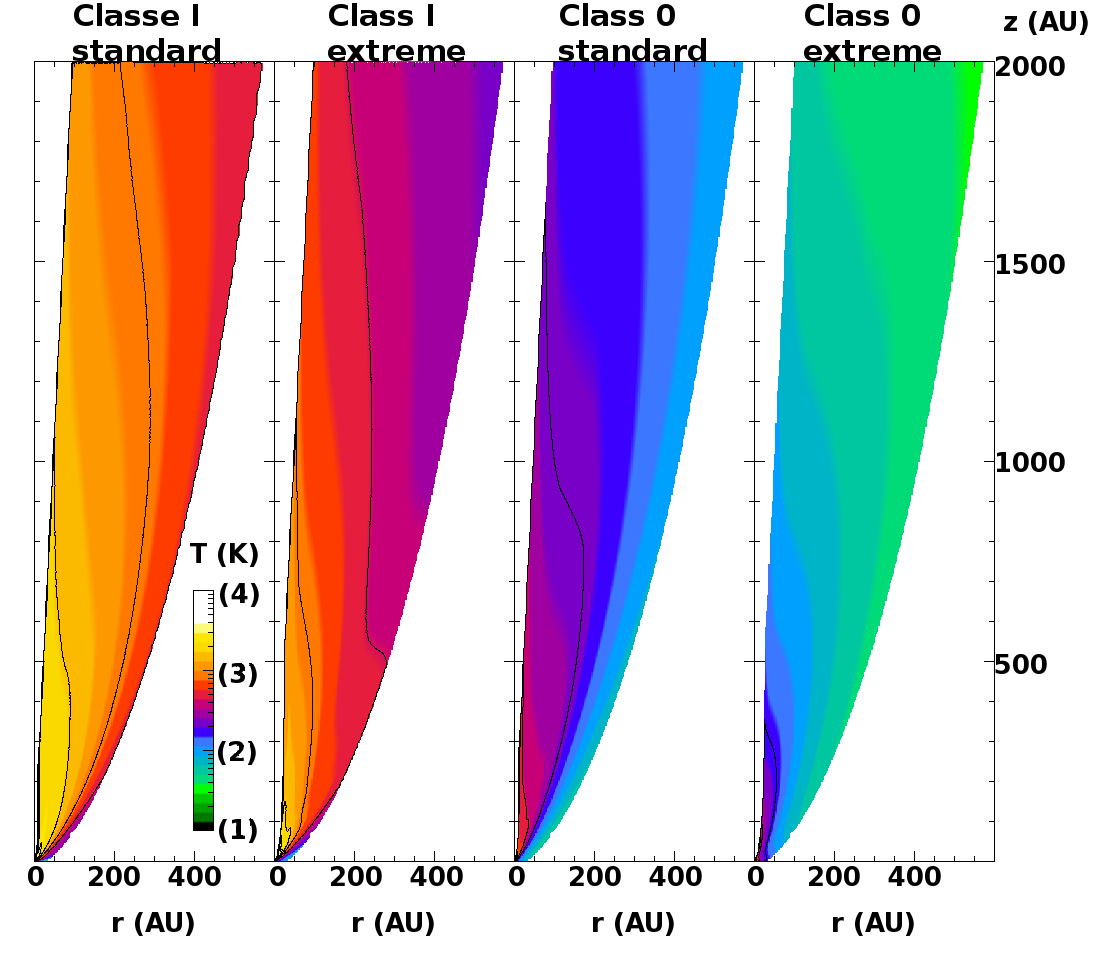}
 \caption{Two-dimensional cuts of the calculated gas kinetic temperature (on a log10 scale) in the four source models with the parameters given in Table~\ref{tab:models}. The black contour levels are at 250, 400, 800, and 1600~K. The streamlines have launch radii ranging from the sublimation radius \Rsub\ to 25 AU. The calculations were stopped inside the cone with $z/r = 20$ where streamlines start to recollimate towards the axis and shocks may later develop. The wind density drops from right to left successively by a factor 4, $\sqrt{5}$, and 5. The wind temperature increases when the wind density drops, as a result of the balance between radiative cooling and heating by ion-neutral drift (caused by MHD wind acceleration). }
\label{fig:temp_cuts}
\end{figure}

%For the \Macc = $ 2.10^{-5} \UMacc$ model
%the kinetic temperature increases by a factor 2-3
%at the sonic point, until $z/r = 1$.
%At $z/r = 1$, \Tkin\ becomes
%lower and reaches factor 2 lower at recollimation of the streamline,
%decreasing monotonously and
%faster along streamlines.

% (dust temperature) --> FIGURE en annexe
% The streamlines launched from $r_0 > 20 \, \AU$ have dust temperature
% $\Tdust \le 100$ $\K$ at
% sonic point.
% The dust temperature become $\le 100$ $\K$ at $R > 30 \, \AU$
% along almost streamlines photodesorption processes are strong enough
% to make ices to desorb.
% for streamlines launched before $r_0 = 6.4 \, \AU$ thermal
% desorption dominates
% adsorption processes at $z/r < 1$ until photodesorption processes takes over.
% But, for the outer streamlines $r_0 > 12.8 \, \AU$ and $z/r < 1$
% our assumption is not valid.
% (H2O heating / cooling small scales) --> COOLING ETAPES
%At small spatial scales $z/r < 0.5$ $\water$ molecules heat the gas
%through IR pumping by diffuse dust background,
%and at $z/r > 0.5$ $\water$ leads the cooling.%, except around
%The dominant heating term is the ambipolar diffusion.

\subsection{Water gas-phase abundance} \label{results_abun_H2O}
%ABUNDANCE

%The observation of ice absorption bands indicates
%that L1527 and Elias 29 have quantity of water ice on grain
%beyond $R = 100\, \AU$ from the star in the envelope
%and without presence of grain ices in the disk \citet{2012AikawaL1527},
%confirming our fully desorbed ices along the jet for this object.

%   \begin{figure}
%   \centering
%   %\includegraphics[width=\linewidth]{FigureA2} %height=18 cm
%   \includegraphics[width=0.9\linewidth]{xH2O_big.png} %FigureA2
%   \caption{Meridional cut of MHD disk wind for the reference model.
%   $Left$ panels~: $\water$ fractional abundance.
%   $Right$ panels~: gas temperature \Tkin\ in $\K$.
%   $Upper$ panel~: jet out to 1500~$\AU$.
%   $Lower$ panel shows zoom centered in inner 100~$\AU$ region. %$\times$ 100~$\AU$ sized.
%   Contours in temperature panel tag isotherm from 100~$\K$ by 100~$\K$
%   in upper panel and 100, 200 to 1000 by 200, 1500
%   and 2000 $\K$ in zoom panel.
%   The inner edge follows the innermost streamline launched at $\Rsub$,
%   streamlines launched from 6.4 and 12.8~$\AU$ are superimposed on
%   $x(\water)$ map in dotted line and the outer edge of the jet follows the
%   streamline launched from $r_0$ = 25~$\AU$. Streamlines starts at $z/r = 0.05$
%   sonic point and end at $z/r = 21$ where recollimation start.}
%   \label{FIGURE1}
%   \end{figure}

Our thermo-chemical model includes 1150 chemical reactions of various types (neutral-neutral and ion-neutral reactions, collisional dissociation, recombinations with electrons, charge exchange with atoms, molecules, grains and PAHs, photodissociation and photoionization by FUV and X-ray photons; see Paper~I), out of which 104 involve gas-phase H$_2$O. However, only three types of reactions are shorter than the wind dynamical timescale and thus play an actual role in the evolution of the wind H$_2$O gas-phase abundance, namely: formation by endo-energetic neutral-neutral reaction of O and OH with H$_2$ (dominant at the base of the wind), formation by dissociative recombination of H$_3$O$^+$ (dominant in the cooler outer wind regions), and photodissociation by FUV radiation emitted by the accretion shock on the stellar surface (attenuated by dust along intervening wind streamlines). The adopted rates for these significant reactions\footnote{Following the referee's request, we have checked the effect of adopting more recent rates for these three types of reactions, which we obtained respectively from a compilation of the NIST Kinetics Database ({\tt kinetics.nist.gov}), \citet{Neau00}, and \citet{vanDishoeck06}. We find that the predicted H$_2$O emissivities are changed by less than 25\% in regions that contribute significantly to emergent line profiles. Our modeling is certainly not accurate down to this level owing to the simplifications made to render the problem tractable (e.g. LVG approximation, 1D dust radiative transfer). Hence, the results obtained with the adopted chemical rates in Table~\ref{tab:H2O-chem} are still adequate for the purpose of this exploratory paper, which is to investigate the overall potential of our MHD disk wind model to explain the observed H$_2$O profiles and correlations with source parameters, not to infer very precise quantities.} are listed in Table~\ref{tab:H2O-chem}. Photodissociation by secondary FUV photons generated by the interaction of stellar X-rays and cosmic rays with molecular gas \citep{Gredel89} is also included in the model, but is too slow to have an effect. Since H$_2$O ice mantles are fully thermally desorbed at the base of the wind for $r_0 \le 25$ AU (see Appendix~\ref{app:initial_abun}), photodesorption reactions by UV and Xray photons --- whose yields as a function of photon energy are still poorly known ---- do not play a role either.

%Re-adsorption on dust grains was neglected, as the timescale was generally longer than the wind dynamical timescale and the opposing reactions of photodesorption by UV and Xray photons are still poorly known. The dominant included processes are summarized, with their adopted rates, in Table~\ref{tab:H2O-chem}. 

\begin{table*} 
\caption{  Chemical reactions affecting the H$_2$O gas-phase abundance in our models}
\label{tab:H2O-chem}
\centering
\begin{tabular}{l l l }
\hline\hline
Reaction        &       Adopted rate per unit volume (${\rm cm^{-3} s^{-1}}$) & Refs. \\
O + H$_2$ $\rightarrow$ OH + H & $1.55\times10^{-13} \left(T_{\rm kin} \over 300{\rm K}\right)^{2.8} \exp(-2480{\rm K}/T_{\rm kin})\, n({\rm O}) \, n({\rm H_2})  $& 1\\
OH + H$_2$ $\rightarrow$ H$_2$O + H  & $9.54\times10^{-13} \left(T_{\rm kin} \over 300{\rm K}\right)^2 \exp(-1490{\rm K}/T_{\rm kin})\, n({\rm OH})\,n({\rm H_2})  $& 1 \\
H$_3$O$^+$ + $e^-$ $\rightarrow$ OH + H$_2$     & $8.45\times10^{-7} \left(T_{\rm kin} \over 300{\rm K}\right)^{-0.5} \,n({\rm H_3O^+})\,n(e^{-}) $ & 2\\
H$_3$O$^+$ + $e^-$ $\rightarrow$ H$_2$O + H     & $4.55\times10^{-7}  \left(T_{\rm kin} \over 300{\rm K}\right)^{-0.5} \,n({\rm H_3O^+})\,n(e^{-}) $ & 2\\
H$_2$O + FUV $\rightarrow$ OH + H       & $8\times10^{-10} \, \chi \exp(-1.7 A_V) \, n({\rm H_2O}) $ & 3\\
H$_2$O + FUV $\rightarrow$ H$_2$O$^+$ + $e^-$   & $3.3\times10^{-11} \, \chi \exp(-3.9A_V) \, n({\rm H_2O}) $ & 3\\
%H$_2$O + X-rays/CR $\rightarrow$ OH + H                                &  $1940 \zeta_X \left({n({\rm H}_2) \over 0.54 n_{\rm H}}\right) \, n({\rm H_2O}) $ & 5,6\\ 
\end{tabular}
\tablefoot{  $n(X)$ is the number density of species $X$ in cm$^{-3}$, $\chi$ is the unattenuated incident stellar FUV field in units of the Draine interstellar FUV field (see Sect. 2.4.4 of Panoglou et al. 2012), $A_V$ is the visual extinction between the star and the local point provided by inner dusty wind streamlines (see Eq.17 of Panoglou et al. 2012). 
%$\zeta_X$ is the total local rate of ionization by cosmic rays and (attenuated) stellar coronal X-rays (see Sect. 2.5.2 of Panoglou et al. 2012). 
Refs: (1) \citet{Cohen83} (2) \citet{Millar89} (3) \citet{vanDishoeck88}.
%(5) \citet{Gredel89}, (6) \citet{FPdF07}.
}
\end{table*}
%\begin{itemize}
%\item Neutral-neutral endothermic reactions 
%\item Photodissociation by the FUV radiation of the accretion hot-spot, attenuated by dust on intervening wind streamlines.
%The adopted 
%\end{itemize}

Figure~\ref{fig:abun_cuts} presents color maps of the calculated 2D distribution of water gas-phase abundance relative to H nuclei in our four source models, obtained by interpolation between the discrete calculated streamlines. We note that the initial gas-phase water abundance at the wind base is $\simeq 10^{-4}$ (see Appendix~\ref{app:initial_abun}), substantially higher than in Paper~I owing to the warmer dust temperature allowing for thermal evaporation of water ice.

\begin{figure}
 \includegraphics[angle=0, width=\linewidth]{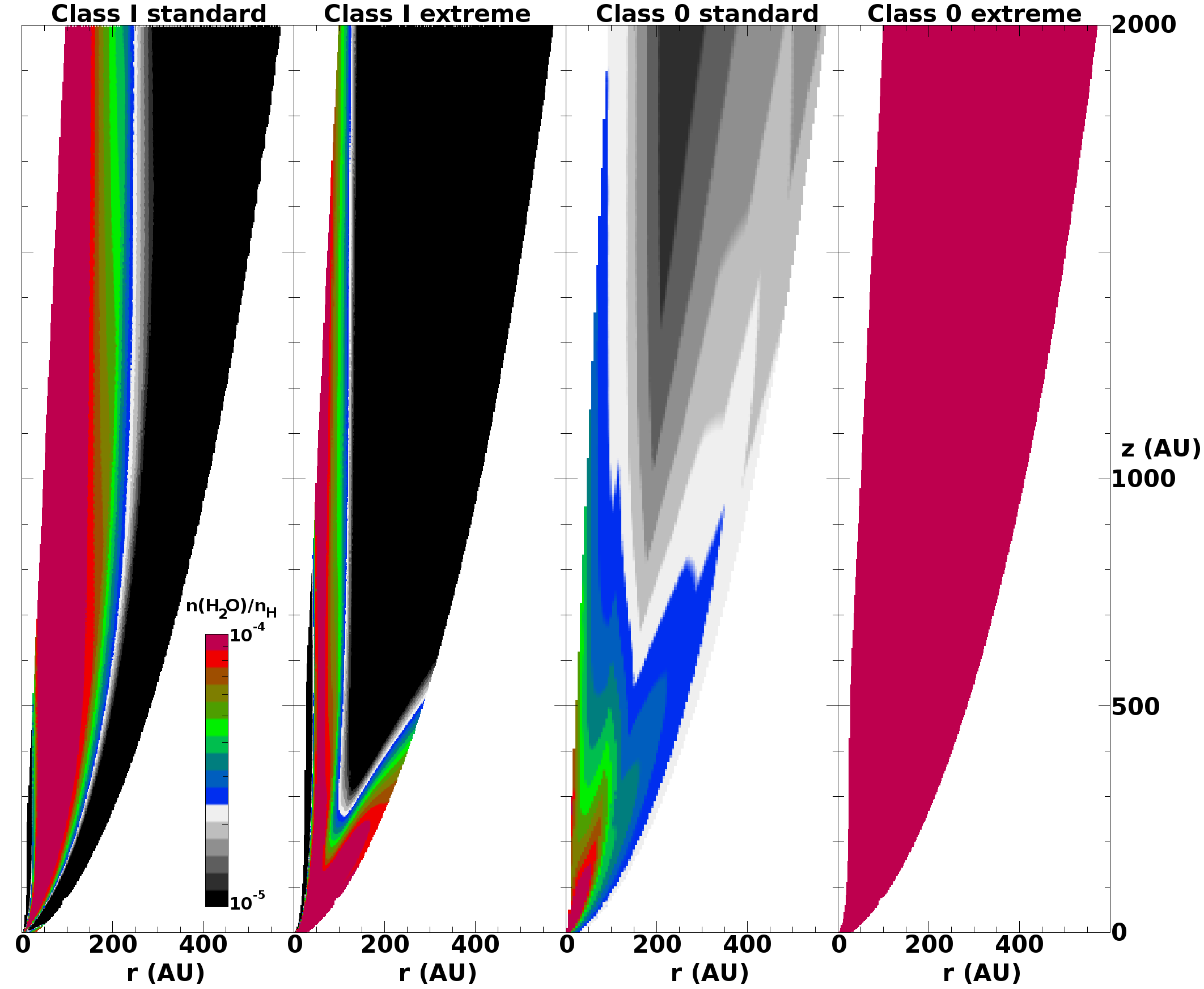}
\caption{Two-dimensional cuts of the calculated H$_2$O gas-phase abundance relative to H nuclei in the four source models with the parameters given in Table~\ref{tab:models} in the same order as in Fig.~\ref{fig:temp_cuts}. The plotted dusty streamlines have launch radii ranging from the dust sublimation radius \Rsub\ to 25 AU. Initial abundances at the wind base are discussed in Appendix~\ref{app:initial_abun}.}
\label{fig:abun_cuts}
\end{figure}

In the densest model (extreme Class 0) the gas-phase water abundance stays constant along all dusty streamlines at the initial value of $\simeq 10^{-4}$  because the disk wind is so dense that its dust content screens it efficiently against stellar FUV and X-rays photons, and water photodissociation is negligible. 
In the other three less dense models, photodissociation starts to operate and the evolution of the water abundance results from a competition between photodissociation and gas-phase reformation through the reaction $\mathrm{OH} + \mathrm{H}_\mathrm{2} \rightarrow \mathrm{H}_\mathrm{2}\mathrm{O} + \mathrm{H}$. Since this reaction is endo-energetic by $\simeq 1500$~K, its speed increases not only with density, but also with temperature as exp(-1500~K/\Tkin).
 
In the standard Class 0 model, FUV photodissociation remains moderate, but gas temperatures reach only a few hundred  K so that reformation is not complete on scales of 2000 AU. The water abundance drops by typically 30\% ($x(\water) \sim 3 \times 10^{-5}$) on this scale. It reaches a minimum $\sim 10^{-5}$ on intermediate streamlines launched from 3--6 AU, which are both less dense (i.e. slower reformation) than the axial regions and less well screened against FUV irradiation than the outermost streamlines.  

In the Class 1 models, the wind is even more transparent than in the S0 model and FUV photodissociation of water is efficient above $z/r \simeq 2$ in the X1 model or right above the wind sonic point in the S1 model. At the same time the wind is warmer than in the S0 model, increasing the rate coefficient for water reformation. This allows  a high water abundance $\sim 10^{-4}$ to be maintained on the axial (densest) streamlines, while the abundance drops to $\le 10^{-6}$ on outer streamlines where density is too low for reformation. Hence Class 1 models have their inner streamlines significantly more water-rich than outer ones, at variance with the shallower gradient predicted for the Class 0. We note that the axial spine of high water abundance extends to larger radii in the standard Class 1 compared to the extreme Class 1: the higher wind temperature in this case (see Fig. \ref{fig:temp_cuts}) makes reformation reactions more competitive than photodissociation. 

\subsection{Synthetic H$_2$O line profiles}

Figure~\ref{fig:profiles} illustrates the influence of our four free parameters (\Mstar, \Macc, \romax, $i$) on the predicted H$_2$O line profiles in 
two lines of ortho-water: the fundamental $1_{10}-1_{01}$ line at 557 GHz and the higher excitation $3_{12}-2_{21}$ line 
at 1153 GHz. {  The 557 GHz line is selected for detailed study because it is the strongest water line, and the only one detected by WISH in Class 1 protostars and so  it will provide our strongest basis for a statistical comparison with observations, including correlations with source properties  (e.g. envelope density).  The 1153 GHz line is among the most excited H$_2$O lines detected with HIFI towards low-mass protostars, with an upper level at $E_{\rm up} \simeq 200$ K above the fundamental level of ortho-water (see  Section~\ref{obs:excited}). We select it to investigate the influence of $E_{\rm up}$ on the emergent line profiles. In the excitation regime matching the observed line ratios in the BC, the excitation conditions probed by low-$J$ water lines are found to depend only on the H$_2$ density and H$_2$O column density, with essentially no sensitivity to temperature above 300~K \citep[see Fig.~11 of][and references therein]{Mottram2014}. In our D-wind models, the regions dense enough to excite the 1153 GHz line are warmer than 300 K, hence the wind density and H$_2$O column  (i.e. abundance) will also be the dominant factors determining the profile shape and intensity of this line, with only a small effect of the gas kinetic temperature}. 

As a {reference case}, in Fig.~\ref{fig:profiles} we adopt the Standard Class 0 model (\Mstar = 0.1 \Msun, \Macc = $5 \times 10^{-6} \, \UMacc$) with \romax = 25 AU and $i = 60$\degr. We then overplot predicted profiles where only one of the four free parameters has been changed.  A typical source distance of $d =235$ pc (Perseus cloud) is adopted. We first present synthetic profiles for only one  lobe of the disk wind (blue), as it makes it is easier to understand the origin of profile features. The result of adding the blue and red lobes is shown in the  rightmost panel (together with the effect of changing inclination).

\noindent

    \begin{figure*} 
    \centering
    %ps2pdf -dDEVICEWIDTHPOINTS=600 -dDEVICEHEIGHTPOINTS=1100 PAPERR0MAX.ps ALLEFFECT.pdf
    \includegraphics[angle=0, width=\linewidth]{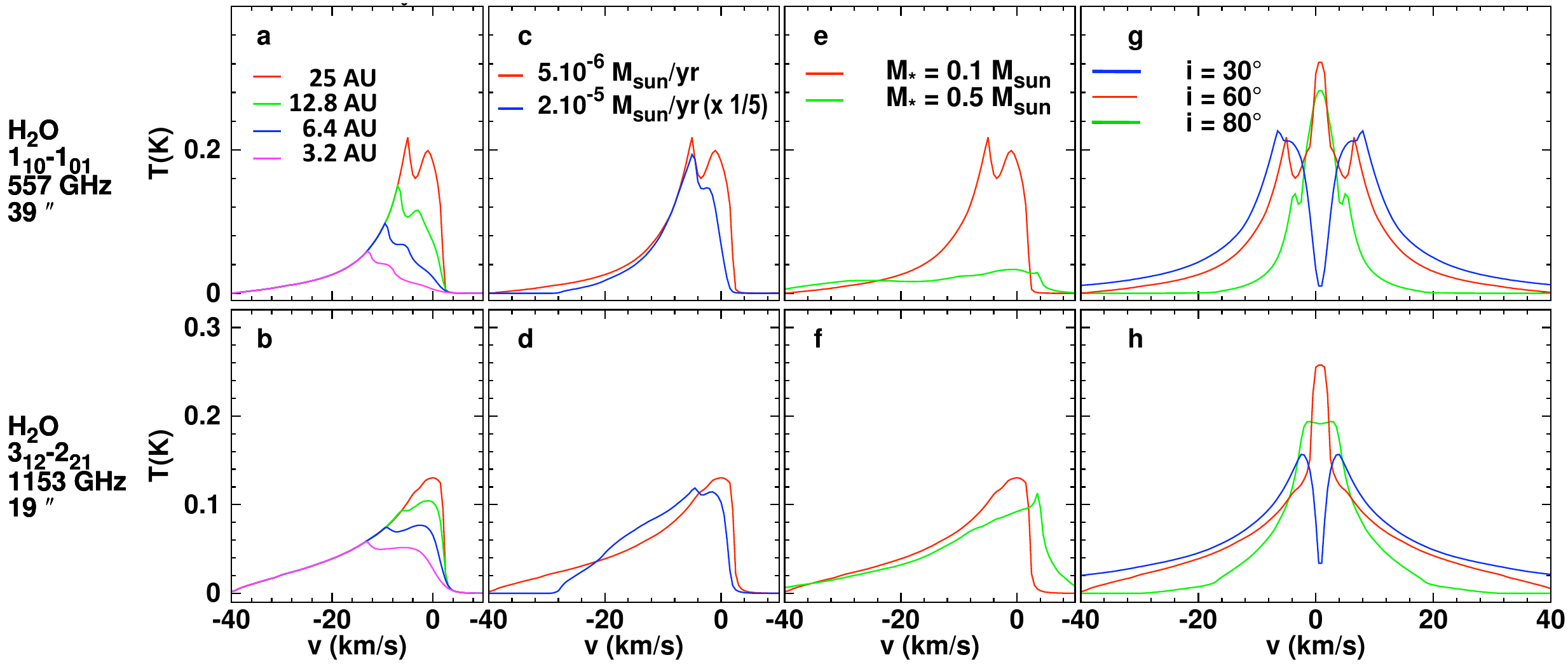}
   \caption{Influence of our four free parameters on synthetic {\em Herschel}/HIFI profiles in the ortho-$\water$ fundamental line $1_{10}-1_{01}$ 
   at 557~$\GHz$ (top row) and the $3_{12}-2_{21}$ line at 1153~$\GHz$ (bottom row). The reference profile (S0 model with \romax = 25 AU and $i = 60$\degr) is plotted in red, and other colors show the effect of changing one of the parameters. From left to right: \romax (panels a--b), accretion rate (panels c--d), \Mstar\ (panels e--f), inclination angle $i$ to the line of sight (panels g--h). Only the blueshifted lobe is shown, except in Panels g-h where red and blue lobes have been summed. The distance to the source is set to 235~$\pc$.  }
    \label{fig:profiles}
    \end{figure*}

%\subsubsection{Generic line profile properties}
%
%The line profile from a single lobe presents two peak,
%a first peak centered on zero and a second higher
%peak at higher velocity.
%%---------------%
%The first peak correspond to
%flux cumulated from a wide region of the jet with
%low velocity along a large range in elevation $z$.
%And the flux seen at positive velocity is emitted from
%regions where rotation and expansion motions are still
%important,
%excitation of high energy levels is significant.
%%---------------%
%The second peak is formed further
%in the accelerated wind region
%at $z/r > 5$
%in range in $z$ as large as for the first peak.
%Flux from regions corresponding to $z/r > 5$
%represents $\sim$30\% of total flux for
%the fundamental lines
%and almost flux for other
%excited lines.
%

\subsubsection{Changing the outermost launch radius \romax\ }
\label{results_r0max}

%The inclination angle is fixed to its reference value of 60\degr.
%The impact of \romax\  is more clearly visible at low inclination angle.
%Varying extension of launching area changes the flux and
%the FWHM through the line peak, but not the line width at zero intensity
%of the lines.
Since the flow speed in our self-similar disk wind model scales as $r_0^{-0.5}$ (Keplerian law), one expects that
the broad range of launch radii contributing to the emission in a disk wind will naturally produce a broad line profile. 
Figure~\ref{fig:profiles} (panels a-b) illustrates how increasing \romax\  from 3.2 to 25~$\AU$ indeed contributes to adding flux at progressively lower and lower velocity because of this $r_0^{-0.5}$ scaling. As a result, a blue sloping wing develops while the red edge of the profile becomes steeper and steeper. 

The quantitative effect of \romax\  on the water line profile depends on the upper energy level. When increasing \romax\  from 3.2 to 25~$\AU$, the integrated line flux increases by a factor of $\sim 2$ in the fundamental 557 GHz line, and by only 30\% for the 1153 GHz line. This occurs because the fundamental line (Panel a) is more sensitive to the contribution of external streamlines than transitions from higher levels such as $3_{12}-2_{21}$ (Panel b), which require higher densities for excitation and are thus dominated by smaller $r_0$. Hence, the fundamental line offers a better diagnostic of \romax\  in the considered range. 

We note that, in contrast to the total line intensity, the full width at zero intensity (FWZI) remains invariant to changes in \romax\  because the FWZI is determined solely by the range of projected velocities along {the fastest, innermost emitting streamline}, here ejected from \Rsub. In the reference case, the line pedestal of the blue jet lobe ranges from about -40 to +4 \kms\ when viewed at 60\degr. The blue edge corresponds to the maximum $z$-velocity (projected on the line of sight) reached along the \Rsub\ streamline, while the red edge is produced by the initial rotation motions near the wind base.   

%The  blue wing in the profile comes from the accelerated upper wind regions at $z/r > 5$ where the gas has reached its maximum speed. The emission at lower velocity comes from regions closer to the wind base at $0.5 < z/r < 5$, where the gas is not yet accelerated and rotation and expansion motions are still important, allowing some slightly redshifted motions on the line of sight \citep[see e.g.][]{Cabrit1999}. This (denser) wind base dominates the flux in the excited line, producing a single peak near systemic velocity, while the upper wind also contributes substantially in the fundamental line, producing a second peak to the blue.

%60\% of the total flux in fundamental line, and 90\% of the flux in excited lines is emitted from $0.5 < z/r < 5$.
%The FWHM at 60\degr decreases from $\sim 30 \, \kms$ to $\sim 10 \kms$
%for the 557~$\GHz$ ortho fundamental line and from $30$ to less than $\sim10 \kms$
%for the 1153~$\GHz$ ortho excited line when \romax\  increases from 6.4 to 25~$\AU$,

% We verified that in all cases , so that the solutions,
% computed under single fluid perfect MHD approximations, remain valid.

%---------------------------------------------------------%
\subsubsection{Changing the accretion rate \Macc} \label{results_macc}

Panels c-d of Figure~\ref{fig:profiles} present the effect of increasing the accretion rate by a factor of 4,
from \Macc = $5 \times 10^{-6} \, \UMacc$ to $2 \times 10^{-5} \, \UMacc$, keeping all other reference parameters fixed. 
This modified set of parameters corresponds to our X0 source model in Table~\ref{tab:models} where
both the wind density and the hot-spot luminosity are multiplied by 4 from the reference S0 model.

The predicted profiles for X0 in Fig.~\ref{fig:profiles}c-d are divided by a factor of 5 to ease comparison with the reference S0 case. It can be seen that the main effect of increasing \Macc\ is to increase the intensity by a factor of $\simeq$5 close to the density increase. 

%This illustrates the fact that most of the line emitting volume is in sub-thermally excited regions well below the critical density, where water is optically thick but "effectively thin"  \citep{KN}. The slight changes in kinetic temperature and water abundance between the two models have a modest (compensating) effect in comparison.

The shape of the profile wings is not strongly affected apart from a slight reduction of the maximum wing extension from $-40$~\kms\ to about $-30$ ~\kms. This results from the four-fold increase in FUV hot-spot luminosity, which increases the sublimation radius \Rsub\ from 0.31 to  0.63~$\AU$ and thus decreases the maximum velocity reached on the innermost \Rsub\ streamline by the ratio of the Keplerian speeds ($\simeq \sqrt{2}$).
%the emission at velocity bluer than -50~\kms\ corresponding to launch radii $r_0 < 0.63 \, \AU$. 
%We explore a 4 times denser disk wind
%with the higher accretion rate 
%(mass and radius of the star stay unchanged),
%see Table~\ref{table_model}. %The dust temperature profile is not greatly modified except close to the sublimation radii, 
%which moves to 0.63~$\AU$. (see Figure~\ref{FIGURE_DUSTYTD} in appendix).

%The increase in density implies an increase in opacity, but
%collisional excitation of $\water$ also increases leading to a net line flux increased by a factor 5 in

%---------------------------------------------------------%
\subsubsection{Changing the stellar mass \Mstar} \label{results_class}

Panels e-f in Figure~\ref{fig:profiles} present the effect of increasing the stellar mass
from 0.1 \Msun\ to 0.5 \Msun, keeping all other reference parameters fixed. 
This modified set of (\Mstar,\Macc) corresponds to our X1 source model in Table~\ref{tab:models}. 
%compare predicted profiles
%for  class 1 sources (with parameters described in Table~\ref{table_model})
%with our reference class 0 model.
%The first class 1 accretion rate $\Macc = 1 \ times 10^{-6} \, \UMacc$
%predicts flux lower by two order of magnitude than class 0 reference model
%but with flux at higher velocity.
%And the higher class 1 accretion rate ,
%and the accretion hot spot luminosity by a factor 5. 
%The FUV excess $= 0.5L_{\rm acc}$ is increased by a factor 5, increasing \Rsub\ from 0.31 to 0.64 AU. At the same time, 
The flow speed along each streamline $V \propto {M_\star}^{0.5}$ increases by a factor of $\simeq 2$, while the wind density $\propto$ \Macc $M_\star^{-0.5}$
drops by the same factor; as a result, the H$_2$O emission from each streamline is shifted to both higher radial velocity and lower intensity. Figure~\ref{fig:profiles}e--f shows that the two effects combine in such a way so as to produce a shape and intensity of the line wings that is very similar to the S0 model. 

Figure~\ref{fig:profiles}e shows that the only clear difference between X1 and S0 profiles appears in the 557 GHz line at projected velocities below -20 \kms\ (for a view angle of 60\degr) where the X1 profile becomes flat-topped. This occurs because the slow outer streamlines that produced the strong central peaks in the X0 profile now become strongly photo-dissociated and water poor in the X1 model (see Fig.~\ref{fig:abun_cuts}). This effect does not occur in the excited line at 1153 GHz as its emission is dominated by the wind base at $z/r < 5$ where H$_2$O is not yet fully photodissociated. We note, however, that at higher inclination $\simeq 80\degr$, the flat-topped part of the 557 GHz line profile in the X1 model will shift to much lower velocities and become difficult to distinguish from the S0 557 GHz profile owing to the unknown envelope contribution at low velocities. This indicates that the fitting of observed profiles with our model grid will have some degeneracy unless there is an independent constraint on \Mstar\ or accretion age (Class) of the source. 

%The similarity in shape and intensity of the broad line wings between the S0 and X1 models means that any  fitting of observations will have some degeneracy unless there is an independent constraint on \Mstar\ or accretion age of the source. 
%\NEW{In both case of accretion rate the zero velocity peak
%corresponds to flux from region located just above the disk plane with $z/r < 0.5$
%and dominated by rotation motion.
%While 60\% of flux in the high velocity peak comes from
%accelerated wind region where $z/r < 5$ and extended up to $z\simeq4000\,\AU$
%and 40\% from region where $z/r > 5$.
%All the flux in the 1153~$\GHz$ line is emitted below $z/r = 5$ in
%limited range region in elevation.}

%---------------------------------------------------------%
\subsubsection{Changing the inclination angle $i$} \label{results_angle}

Panels g-h in Figure~\ref{fig:profiles} present the effect of varying the inclination angle $i$, with all other parameters set to their reference value.
%We have chosen three values for the angle~: near edge-on (80\degr),
%median (60\degr, reference model) and near pole-on (30\degr).
%The \romax\  parameter is fixed to reference value of 25~$\AU$ (see table~\ref{table_model}).
%The reference model at $i = 60\degr$ is shown in red. 
In order to prepare for the comparison with {\em Herschel}/HIFI observations in Section~\ref{sec:obs}, we show here the ``double sided'' profiles obtained by summing the emission from the blueshifted and redshifted lobes of the disk wind. 

In doing this summation we neglected radiative coupling between the two lobes (i.e. absorption of the red lobe by the blue lobe in front of it), the rationale being that at low inclinations $\le 30\degr$ the two lobes are well separated in velocity and thus cannot absorb each other's line emission, while at high inclinations $\ge 80\degr$ the two lobes are projected sufficiently well apart on the sky that they should suffer minimal overlap on each individual line of sight. This simplifying assumption may not be fully adequate at intermediate inclinations $\simeq 60\degr$. However, we stress that our main uncertainty in H$_2$O profile predictions at low velocity is not the blue/red overlap, but the unknown envelope contribution (in emission or absorption) whose complex modeling lies well outside the scope of the present paper \citep[see e.g.][]{Schmalzl2014}.  Our comparison with observed HIFI line profiles will thus focus on the broad wing emission beyond $\simeq$ 5\kms\ from systemic, where the envelope contribution and blue/red overlaps should be minimal.

Figures~\ref{fig:profiles}g--h show that the main effect of increasing the inclination angle $i$ is to reduce the base width of the line profile. This results from the cos$i$ term involved in the projection of the maximum jet speed directed along the $z$-axis. The line width at zero intensity shrinks from  $\pm$ 60~\kms\ at 30\degr\ to $\pm$20~\kms\ at $i = 80\degr$. This effect will allow us to constrain the inclination angle when comparing it with observed line profiles. 

Another effect of observing farther away from pole-on is that it decreases the LVG line escape probability towards the line of sight, $\beta_\mathrm{los}$, because of the smaller projected velocity gradient when looking at the jet sideways rather than down its axis. When the inclination increases from 30\degr\ to 80\degr, the integrated line intensity, \TdV, drops by a factor of 2.2 for the fundamental line at 557~$\GHz$ and by a factor of 1.7 for the excited line at 1153~$\GHz$.

\subsection{Scaling of \TdV\ with distance}
\label{TdV_distance}

In order to compare the intrinsic water line luminosity of protostellar objects observed at various distances $d$, it is often assumed that the observed \TdV\ scales as $d^{-2}$ \citep[e.g.][]{Kristensen2012557GHz}, meaning that the line flux comes mainly from an unresolved region. However, if the emission is unresolved across the flow axis but resolved and uniformly bright {along} the flow axis, \TdV\ will instead scale as $d^{-1}$ \citep[e.g.][]{Mottram2014}. We have thus investigated which scaling holds for our disk wind models, 
where the emission is centrally peaked but still somewhat extended. 

We have calculated  $T \dr V$ for the 557~$\GHz$ and 1153~$\GHz$ lines in the Herschel beam for a range of source distances $d <$100 to 400~$\pc$
and then fitted the results as a power law $d^{-\alpha}$. 
We find that the value of $\alpha$ depends mainly on the line excitation energy, the wind density, and the distance range, while \romax\  and inclination have a smaller effect. 
The largest deviation from the point-source scaling is encountered in the fundamental 557 GHz line and 
for our highest value of \Macc = $2 \times 10^{-5}$ (extreme Class 0 model), where the outer wind regions begin to contribute significantly. These regions start to be resolved at $d < 100$ pc where our fit yields $\alpha \simeq 1.3-1.4$. However, for all larger distances $d > 100$ pc or smaller \Macc, we find that $\alpha \simeq 1.7-2$. Lines from higher energy levels like $3_{12}-2_{21}$
are dominated by denser and more compact jet regions than the fundamental line, and are thus seen as even more point-like with an $\alpha$ close to 2. 

We conclude that if our disk wind model is adequate, using $\alpha = 2$ is an acceptable approximation in most cases for the purpose of comparing the 557 GHz intrinsic brightness of sources at various distances and looking for global correlations with other parameters. Nevertheless, to determine the best-fitting grid parameters for each observed source, we will recompute synthetic line profiles at the corresponding source distance to minimize systematic effects. 

%For large distances $d \ge 200$ pc, we find that $\alpha \simeq 2$,
%the whole emitting jet is covered by the beam and seen as point-like.
%While at smaller distance
%($100 \, \pc < d < 220 \, \pc$ for the 557~$\GHz$ line and $d < 180 \, \pc$ for the
%1153~$\GHz$ line)
%$\alpha \simeq -1.6 $ to $-1.8$ ().
%Only streamlines launched from $r_0 < 10 \AU$ are resolved
%regardless to the rest frequency.
%Small effect on alpha (order 2): Near pole-on ($i = 30$\degr) emitting part of the jet is much
%largely included in the beam and produce higher values of $\alpha$
%\citet[appendix B]{Tafalla2010}.

%--------------------------------------------------------------------------------------------------%
\section{Comparison with {\em Herschel}/HIFI H$_2$O observations of protostars}\label{sec:obs}
%--------------------------------------------------------------------------------------------------%

We now compare our model predictions with the results of H$_2$O rotational line profile observations 
towards 29 low-mass Class 0 and Class 1 protostars obtained with the HIFI spectrometer 
on board {\em Herschel} in the framework of the WISH Key Program
\citep{WISH}. We explore whether our grid of dusty MHD disk wind models can reproduce 
the shape and intensity of the broad component of  the fundamental H$_2$O line at 557 GHz, as well as the correlation of its integrated intensity with envelope density found by \citet{Kristensen2012557GHz}. We also investigate whether an MHD disk wind model can explain the relative intensities of more excited H$_2$O lines.

\subsection{The 557 GHz broad component profile}\label{sec:557-profiles}

\noindent

   \begin{figure*}%[h!]    \centering
   \includegraphics[width=0.9\textwidth]{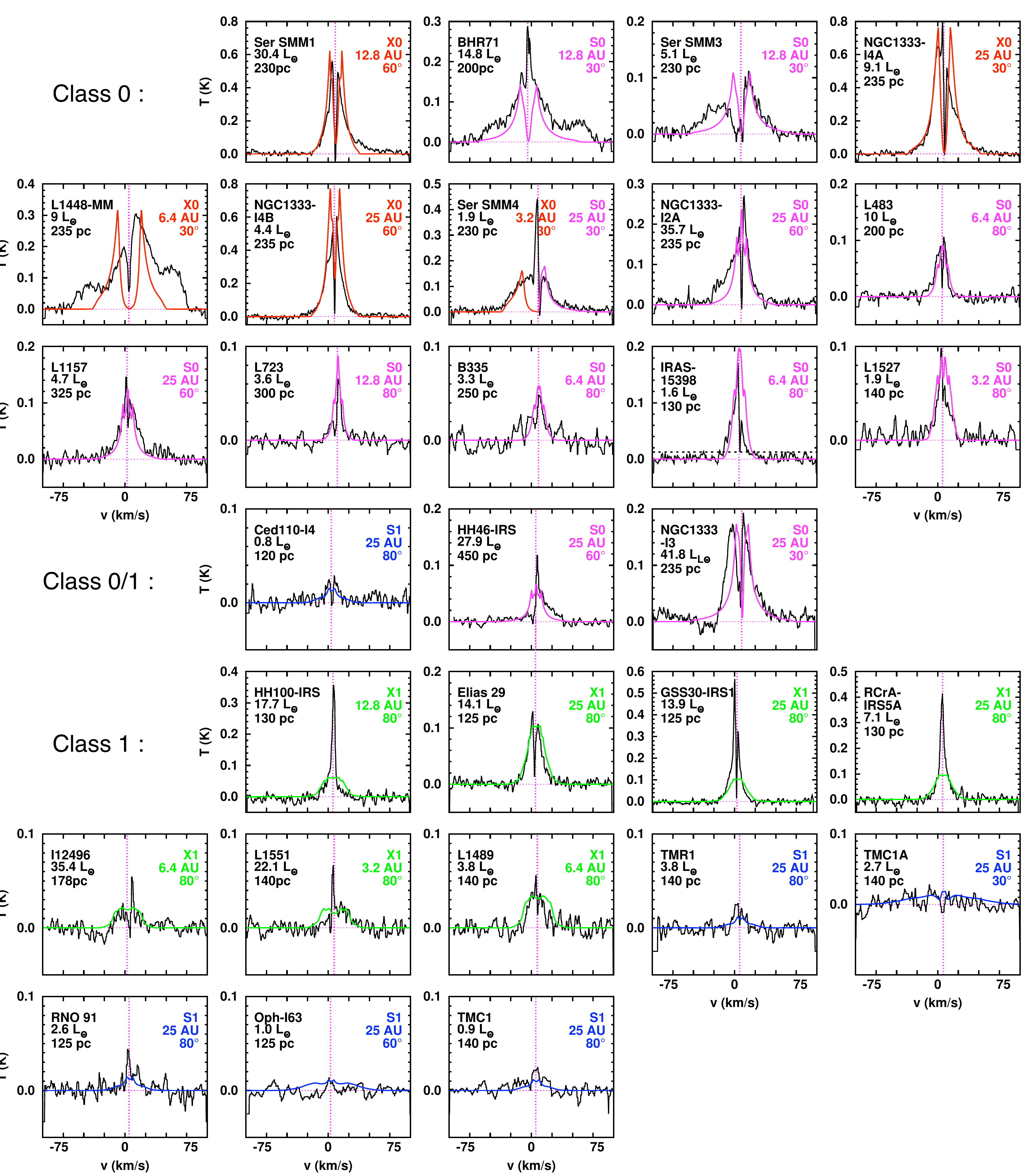}\vspace*{-1mm}
   \caption{Line profiles of the o-H$_2$O fundamental line at 557 $\GHz$ towards 29 Class 0 and Class 1 low-mass protostars as observed with Herschel/HIFI (in $black$, from \citet{Kristensen2012557GHz}), superimposed with the closest matching line profile from our grid of 48 models ($red$ = X0 model, $pink$ = S0, $green$ = X1, $blue$ = S1).
   Predicted profiles were calculated for the specific distance and rest velocity of each source without any arbitrary intensity rescaling.
   The model parameters (source model, \romax, and $i$) are written in the top right corner of each panel, and are summarized in Table~\ref{tab:fit}.
   The plotted model is of the same  evolutionary Class as that attributed by \citet{Kristensen2012557GHz} on the basis of \Tbol, except for three borderline sources denoted ``Class 0/1''  (see text). In Ser MM4, two different models are shown for the blue and red wings (X0 and S0, respectively). 
   }
   %The last six plots show class 1 sources from \citet{Kristensen2012557GHz}
   %which are possibly fitted by class 0 models.}
   \label{fig:BestFit}
\vspace*{-1mm}
   \end{figure*}

\noindent

%-----
\begin{table*}
\caption{Grid models that best match the entire 557 $\GHz$ broad line wings}
\label{tab:fit}
\centering
%\begin{tabular}{ p{3 cm} | p{1.2 cm} p{1.2 cm} p{1.2 cm} p{1.5 cm}p{1.0 cm} p{1.5 cm}}
\begin{tabular}{ l l c c l l l}
%\begin{tabular}{p p p{1.5 cm}          p{1.0 cm} p{1.5 cm}}
\hline\hline
             Source                          &
             $d$ ($\pc$)                     &
             $L_\mathrm{bol}^a$ ($\Lsun$)      &
              \romax\  ($\AU$)               & 
             $i_\mathrm{fit}$ (\degr)        &
             $i_\mathrm{obs}$ (\degr)                   &
             Ref. \\
\hline    %dist
%  Class 0           &       &         &    &             &     \smallskip \\ 
\multicolumn{7}{c}{X0 model: \Mstar = 0.1\Msun, \Macc = $2 \times 10^{-5}$ $\UMacc$} \\
\hline
Ser SMM1                & 230   & 30.4          & 12.8    & 60      \\
NGC1333-IRAS~4A  & 235   & 9.1          & 25            & 30        \\
L1448-MM                & 235   & 9             &  6.4          & 30   & 69                      & 1 \\
NGC1333-IRAS~4B  & 235   & 4.4          &  25   & 60   & 50--80                 & 2 \\
Ser SMM4 (blue)         & 230   & 1.9           & 3.2   & 30     \\
\hline
\multicolumn{7}{c}{S0 model: \Mstar = 0.1\Msun, \Macc = $5 \times 10^{-6}$ $\UMacc$} \\
\hline
BHR71                           & 200   & 14.8   & 12.8 & 30       \\
Ser SMM3                & 230   & 5.1     & 12.8        & 30      \\
Ser SMM4 (red)          & 230   & 1.9     & 25          & 30     \\
NGC1333-IRAS2A          & 235   & 35.7   & 25           & 60      \\
L483                            & 200   & 10      & 6.4 & 80     \\
L1157                           & 325   & 4.7     & 25          & 60  & $80 \pm 15$                 &  3 \\
L723                            & 300   & 3.6     & 12.8        & 80      \\
B335                            & 250   & 3.3     & 6.4 & 80  &  $85 \pm 3$              &  4 \\
IRAS~15398              & 130   & 1.6     &  6.4        & 80  & $\ge 70$                         &  5  \\
L1527                           & 140   & 1.9     & 3.2 & 80  & 85                                      &  6 \\
{\em HH46-IRS}        & 450   & 27.9  & 25              & 60  & $52.5 \pm 2.5$            &  7 \\
{\em NGC1333-IRAS~3(SVS13)} & 235 & 41.8 &  25  & 30  & $26\pm10$       &  8  \\
\hline
%  Class 1           &       &         &    &             &        \smallskip \\
\multicolumn{7}{c}{X1 model: \Mstar = 0.5\Msun, \Macc = $5 \times 10^{-6}$ $\UMacc$} \\
\hline
HH100-IRS               & 130   & 17.7    & 12.8   & 80                 \\
Elias 29                        & 125   & 14.1    & 25  & 80         \\
GSS30-IRS1              & 125   & 13.9    & 25  & 80  & $62.5\pm2.5$            & 9  \\
RCrA--IRS5A             & 130   & 7.1           & 25    & 80            \\
IRAS~12496              & 178   & 35.4    & 6.4         & 80    \\
L1551-IRS5              & 140   & 22.1  & 3.2   & 80  & $45\pm10$                       & 10 \\
L1489                           & 140   & 3.8           & 6.4   & 80  & $74\pm17$                         & 11 \\
\hline
\multicolumn{7}{c}{S1 model: \Mstar = 0.5\Msun, \Macc = $10^{-6}$ $\UMacc$} \\
\hline
{\em Ced110-IRS4}& 120          & 0.8           &  25  & 80     \\
TMR1                            & 140   & 3.8           &  25  & 80   \\
TMC1A                           & 140           & 2.7     &  25  & 30   \\
RNO~91                          & 125           & 2.6     &  25  & 80  & $70\pm4$                &12  \\
Oph-IRS63               & 125           & 1.0     &  25  & 60    \\
TMC1                            & 140           & 0.9     &  25  & 80   
\end{tabular}
\tablefoot{Sources in same order as in Figure~\ref{fig:BestFit}, except for the three borderline Class 0/1 sources in italics (see text). 
$^{a}$Values of $d$ and \Lbol\ from \citet{Kristensen2012557GHz}. References for independent inclination estimates $i_\mathrm{obs}$ (measured with respect to the line of sight): 
(1)\citet{L1448-SiOppm}, % Girart & Acord 2001 find 21deg from plane of sky from SiO jet proper motion
(2)\citet{Desmurs09}, % IRAS4B H2O maser proper motions
(3)\citet{Gueth96}, % L1157-incl
(4)\citet{Cabrit88}, % B335 incl 
(5) This work (Footnote~\ref{foot:iras15398}), % IRAS15398 
(6)\citet{2008TobinL1527}, %L1527 incl (Tobin 2010 takes i = 85 as fixed,referring to their 2008 paper) 
(7)\citet{HH46-incl}, % Hartigan et al 2005 HST proper motion
(8)\citet{HH7-11-incl}, % SVS13: Noriega-Crespo and Garnavich 2001: 
(9)\citet{GSS30-incl}, %Chrysostomou et al 1996
(10)\citet{L1551-incl}, %Hartigan2000 FP obs+bow modeling: i = 35-55deg
(11)\citet{Brinch2007a}, %L1489: 74deg+16/-17=57-90 (molec+scatt light); Padgett et al 1999 i > 60deg from scattered light modeling; 
(12)\citet{RNO91-Lee2000}. % Lee et al 2000 ApJ 542, 925  CO cavity modeling
%\citet{Harsono} % TMC1, TMC1A, TMR1 disk inclinations from dust continuum uv-fit: too loosely constrained
}
\end{table*}

Figure~\ref{fig:BestFit} compares the profile of the ortho-$\water$ fundamental line $1_{10}-1_{01}$ at 557~$\GHz$ observed with {\em Herschel}/HIFI towards 29 low-luminosity Class 0 and Class 1 protostars by \citet{Kristensen2012557GHz} 
with the closest matching model among our grid of 48 cases. Adjustments were made on the broad wings of the central line component, excluding emission within $\pm 5$ \kms\ from systemic where the envelope contribution becomes significant. Each model was recomputed at the specific distance of each source and convolved by the 39\arcsec\ HIFI beam.  {No arbitrary rescaling of the model intensity was applied.} 

When both the S0 and X1 models gave equally acceptable adjustments (cf. Fig.~\ref{fig:profiles}e), we favored the  model consistent with the empirical source classification (Class 0 or Class 1) from the SED bolometric temperature $T_{\rm bol}$ as quoted by \citet{Kristensen2012557GHz}. Figure~\ref{fig:BestFit} shows that a dusty MHD disk wind with values of \Macc\ and \Mstar\ typical of low-mass protostars, accretion age consistent with the source Class, and maximum launch radius \romax\ in the range 3--25 $\AU$ can simultaneously reproduce to an impressive degree {the shape, width, and intensity of at least one wing (and often both)  of the broad central component of the 557 GHz line} in all observed sources, except L1448-MM where the line wings are unusually broad\footnote{We  see in Section~\ref{sec:L1448} that the relative peak intensities among excited H$_2$O lines in L1448-MM are still  reproduced well by our model.}. This overall agreement between model and observations is especially noteworthy considering our use of only one MHD solution, and our sparse parameter grid.

 Given our coarse grid, we note that the model adjustments in Figure~\ref{fig:BestFit}  should only be taken as illustrative examples of the ability of MHD disk winds to explain the broad component of the 557 GHz line in protostars and not as accurate determinations of \Mstar, \Macc, $i$, and \romax\ in each specific source. For example, the similarity between the S0 and X1 line wings seen in Fig.~\ref{fig:profiles}e demonstrates that there can be some degeneracy and that different parameter combinations could probably reproduce equally well or better the line profile in some cases. 
In the present exploratory study, we thus focus on the  general trends of the model and do not attempt to refine the grid to provide optimal fits to each specific source.  We also note that the value of \romax\ used for the models plotted in Figure~\ref{fig:BestFit}  is the {maximum} in our grid that is compatible with the observed line shape, but the true \romax\ could be smaller if other physical components contribute to the inner line wings (e.g. a swept-up outflow cavity, reverse shock fronts in the wind). Table~\ref{tab:fit} summarizes the parameters of the plotted grid model for each source. 

In three sources (Ced110-IRAS4, HH46-IRS, and NGC1333-IRAS3 = SVS13) we were able to find a good fit only with a model of a different Class than indicated by their $T_{\rm bol}$ value. Their spectra are presented under the label ``Class 0 / 1'' in Figure~\ref{fig:BestFit}. We note that the same three sources stand out in the sample of \citet{Kristensen2012557GHz} as exhibiting envelope properties more typical of the other Class: HH46 and SVS13 have large envelop masses typical of Class 0 sources, while conversely Ced110-IRAS4 has the smallest envelope mass in the whole sample, smaller than all the sources classified as Class 1. Hence, these three sources may be borderline cases where the $T_{\rm bol}$ classification is ambiguous. For example, using the ratio of submm to bolometric luminosity as an alternative criterion would make Ced110-IRAS4 change from Class 0 to Class 1 \citep{2001LehtinenCed110,AndreClass0Class1}, in line with our modeling of its H$_2$O line profile. 

We did not attempt to reproduce the extremely high-velocity (EHV) ``bullets'' seen as discrete secondary peaks at $\pm 25-50$\kms\ on either side of the central broad component in a few sources (BHR71, L1448-MM, and L1157). High-resolution studies of similar EHV bullets in SiO and CO have shown that they probe discrete internal shocks along the jet axis \citep[e.g.][]{Guilloteau1992,I04166-PdBI}. Our steady disk wind model does not include shocks and thus cannot reproduce these features. The contribution of shocks might also explain why some sources show slight excess emission in one wing compared to our symmetric steady model, such as  the blue wing in NGC 1333-IRAS 2A (see discussion  in Sect.~\ref{sec:NGC1333}) and in Ser~MM3. Alternatively, it may be possible that the ejection process is intrinsically asymmetric. Ser~MM4 is a case in point, where the red and blue wings may be reproduced by disk wind models viewed at the same inclination and with the same stellar mass but with a different accretion rate and \romax\ in either lobe.  

\subsection{Line profiles of excited H$_2$O lines}
\label{obs:excited}

Another important  test of the MHD disk wind model as a candidate for the broad component of the H$_2$O 557 GHz line in protostars
is whether it could also explain the component shape and intensity in lines of higher excitation. Sources from Figure~\ref{fig:BestFit} with a rich dataset in this respect are the three Class 0 protostars in NGC 1333, and L1448-MM. We examine each of these sets in turn.

\subsubsection{Class 0 protostars in NGC~1333}
\label{sec:NGC1333}

% \noindent
%----ok
NGC1333 IRAS~2A, IRAS~4A, and IRAS~4B are three well-studied prototypes of 
embedded Class 0 protostars in the Perseus complex ($d = 235\pc$).
They were observed with {\em Herschel}/HIFI in several water lines by
\citet{Kristensen2010IRAS}. Figure 5 presents a comparison
of all observed $\water$ line profiles  in each source with synthetic predictions 
for our grid models that most closely match the fundamental 557~$\GHz$ profile (reproduced in the top row). 
The same models with a smaller \romax\  are also shown as dashed curves.

   \begin{figure}
   \label{fig:perseus}
   \centering
      %ps2pdf -dDEVICEWIDTHPOINTS=600 -dDEVICEHEIGHTPOINTS=900  IRAS2A.ps IRAS.pdf
      \includegraphics[width=\linewidth]{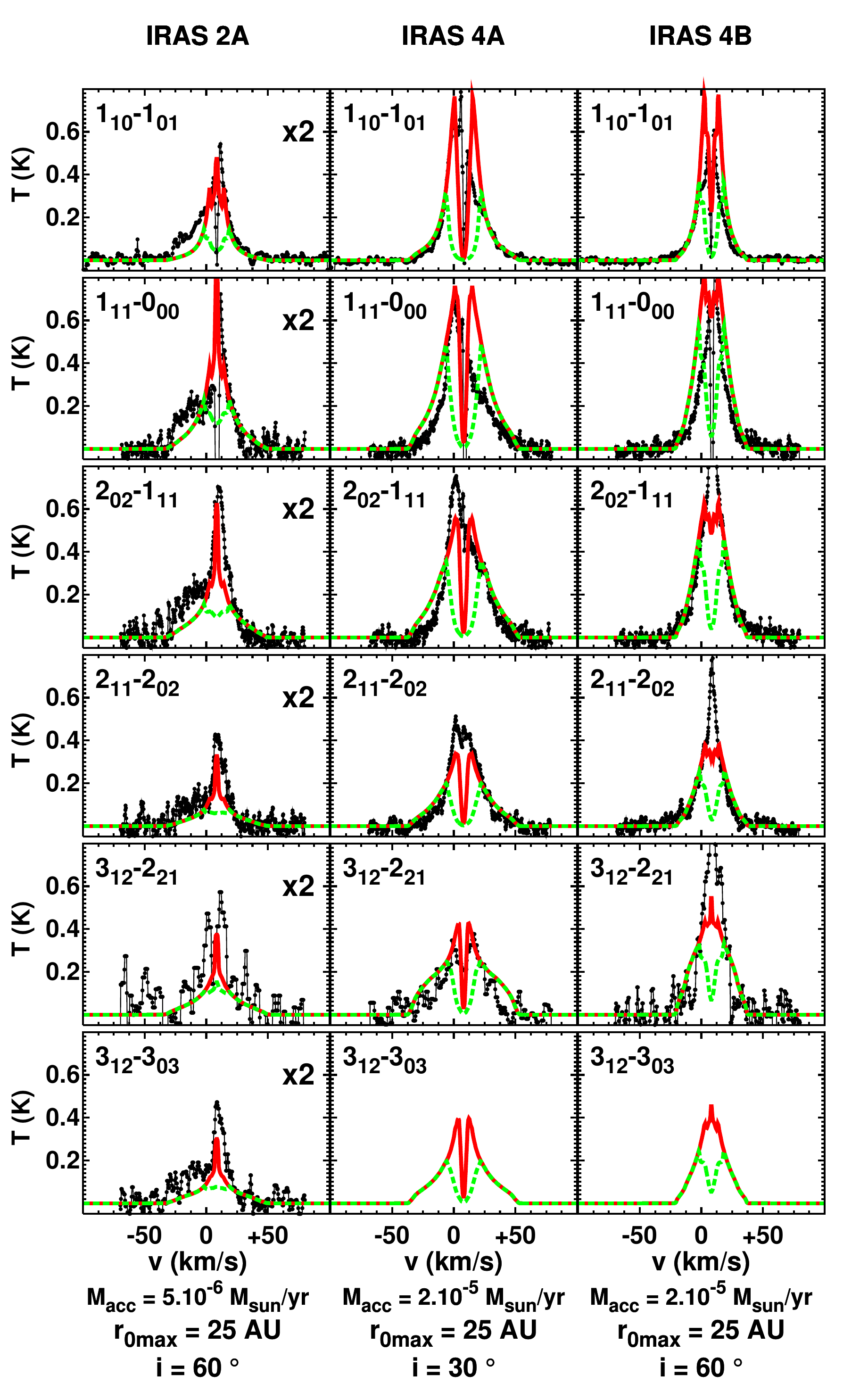}     %IRAS_scaledBESTFIT
     \caption{Comparison of observed $\water$ line profiles (in black) towards NGC1333-IRAS 2A, 4A, and 4B 
     by \citet{Kristensen2010IRAS} with synthetic predictions (solid red curves) for the grid model best fitting the 557~$\GHz$ line in each source with  parameters listed at the bottom. Predicted profiles for a smaller \romax = 6.4~$\AU$ are also shown for comparison (dashed green). }
   \end{figure}

It can be seen that overall, the grid model that best matches the broad component of the 557 GHz line also reproduces remarkably well the wing intensity and shape in more excited H$_2$O lines. 
%(except for a slightly overestimated width in the $1_{11}-0_{00}$ and $3_{12} - 2_{21}$ lines in IRAS 4A/B). 
Therefore, the excitation conditions in molecular magneto-centrifugal disk winds seem adequate to explain the broad component in excited H$_2$O lines as well. The asymmetric blue wing in IRAS 2A, present in all lines, would trace an additional emission component not included in the steady disk wind model, e.g. shocks. The contribution of shocks to the blue wing H$_2$O emission in IRAS2A is supported by the detection in the same velocity range of dissociative shock tracers in {\it Herschel} spectra \citep{Jshocks-Kristensen} and of compact bullets along the jet in the interferometric map of \citet{IRAS2A-Codella}.

%The red-shifted wing is also well
%fitted for the whole lines presented,
%but the model unpredicts the blue wing.
%
%For IRAS 4A, the predicted water profiles
%at 557~$\GHz$ reproduce very well the
%red and blue-shifted wings for the fundamentals
%and $2_{11}-2_{02}$ lines,
%and .
%Because our model does not include the contribution
%of infalling molecular envelope,
%our synthetic predictions are
%not expected to reproduce the inverse P-Cygni
%and red absorption seen in some lines.
%
%IRAS 4B is fitted with same accretion rate and same \romax\ 
%as IRAS 4B but seen under 60\degr instead of 30\degr.
%Line shapes and relative intensities are again
%compatible with our model,
%although $p$-$\water$ fundamental
%synthetic profile is a bit too broad.

\subsubsection{L1448-MM}
\label{sec:L1448}

   \begin{figure}
   \label{fig:L1448-profiles} 
   \centering
     \includegraphics[width=\linewidth]{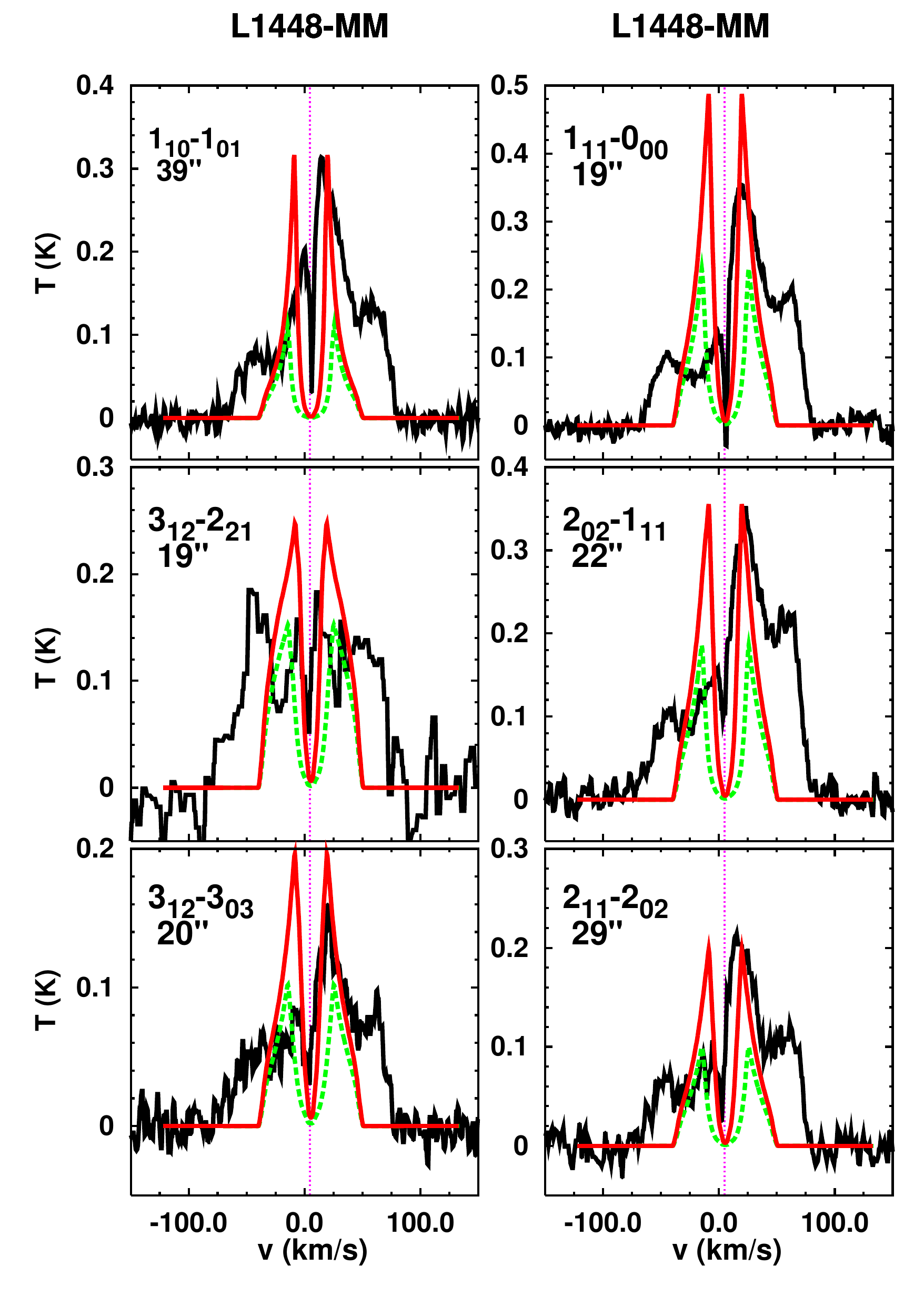}
          \caption{Observed $\water$ line profiles towards L1448-MM 
     \citep{Kristensen2011L1448MM} compared with synthetic predictions (in color) for the grid model best fitting the 557~$\GHz$ line (reproduced in the top left panel). Red solid curves are for \romax = 6.4~$\AU$ and green dashed curves are for \romax = 3.2~$\AU$. The former fits  the red wing intensities better, while the latter fits  the blue wings better. The high-velocity bullets at $\pm 50$\kms\ trace axial shocks not included in our steady-state wind model.}
   \end{figure}

L1448-MM in Perseus ($d = 235\pc$) is the first source where high-velocity bullets were detected \citep{bullets-Bachiller1990}, and where source characteristics were identified as pointing to an earlier evolutionary status than Class 1 sources \citep{L1448MM-Bachiller1991}, eventually leading to the introduction of the new Class 0 type \citep{Class0term-Andre1993}. Figure~6 compares the six water line profiles observed  with {\em Herschel}/HIFI towards L1448-MM by
\citet{Kristensen2011L1448MM} with predictions for the grid model best matching the 557~$GHz$ line profile. As noted earlier, we do not attempt to reproduce the separate peaks centered at $\pm 50$ \kms\ (EHV bullets), which are known to trace internal shocks not included in our steady-state model, and we concentrate on the central broad component. While the modeled 557 GHz line wings are too narrow, the situation improves for lines of higher excitation. The evolution in relative peak intensities is also  reproduced well by the model. 

Figure~6 also shows that the weaker blue wing could be reproduced by a smaller \romax = 3.2 AU than the red wing, where \romax = 6.4 AU gives a good match. Alternatively, the intensity asymmetry between the blue and red wings could reflect different H$_2$O abundances between the two lobes,  caused for example by a different level of external FUV illumination. A possible extreme example of this effect is HH46-IRS, where the blue jet lobe is known to propagate out of the parent globule into a diffuse HII region, and the blue wing of the H$_2$O 557 GHz profile is almost entirely lacking (cf. Fig~\ref{fig:BestFit}).

%Disk-wind launch conditions that differ between the two faces of the disk are conceivable and have been recently simulated numerically (Fendt two-sided).   

%not reproduced by the model.

%The last column of Table~\ref{tableNGC1333} compares the predicted $T \dr V$ in the model and in the observed broad central component  of L1448-MM (as estimated from a Gaussian fitting decomposition by ??). {  The Gaussian fit corrects for central absorption and thus should always give higher TdV than the model ? for the model we should give average of 3.2 and 6.4 AU since each lobe is fitted by a different \romax\  ?}.

%\subsubsection{Ser-SMM1}\label{observations_SMM1}

%{  Show here the excitation diagram for the PACS lines in SMM1 vs the model ?}

\subsection{Correlation of 557 GHz line luminosity with envelope density} 
%\label{sec:TdV-source}
\label{sec:TdV-nH}

%\citet{Kristensen2012557GHz} found significant correlations in their sample between $TdV$(200 pc), the integrated intensity of the 557 GHz line in the {\em Herschel} beam scaled to a distance of 200 pc, and the envelope density and bolometric luminosity of the protostellar source. An important further test of the disk-wind hypothesis is whether it can reproduce these global correlations, and the locus that Class 0 and Class 1 sources occupy in them. 

%In Figures~\ref{fig:TdV-nH} and ~\ref{fig:TdV-Lbol} we overplot the predictions of our 4 disk wind models (colored rectangles) on top of these observed correlations using data from \citet{Kristensen2012557GHz}. The vertical range of $TdV$(200 pc) for each model illustrates the effect of varying \romax\ between 3.6 and 25 AU and the inclination between 80\degr and 30\degr. We discuss each correlation in turn below. 

%\subsection{Envelope density}
%\label{sec:TdV-nH}

\citet{Kristensen2012557GHz} searched for correlations in their sample between \TdV(200 pc), the integrated intensity of the 557 GHz line in the {\em Herschel} beam scaled to a distance of 200 pc (i.e. a quantity proportional to the line luminosity) and various intrinsic properties of the source. The tightest empirical correlation they obtained was with the envelope density at 1000 AU as determined by fitting the integrated source SED with the DUSTY 1D code. 
%It is thus an important test of the disk-wind hypothesis to check whether it can reproduce this correlation. 
It is thus important to check whether our MHD disk-wind model could explain this good correlation and its slope close to 1. We note that the quantity denoted \nH(1000 AU) in \citet{Kristensen2012557GHz} was actually the number density of H$_2$ molecules (Kristensen, private communication). To avoid any ambiguity, here we use  the notation \nenv(1000 AU). 

In Figure~\ref{fig:TdV-nH} we plot as boxes the predicted locus of our four source models in the \TdV(200 pc) -- \nenv(1000 AU) plane.  The datapoints from \citet{Kristensen2012557GHz} are superimposed for comparison. 
The vertical range in \TdV(200 pc) for each model illustrates the effect of varying \romax\ between 3.6 and 25 AU and the inclination between 80\degr and 30\degr. The horizontal range in envelope density plotted for each model is 1--3 times \nff(1000 AU) (the fiducial envelope density given in Col. 4 of Table~\ref{tab:models}), i.e. it assumes an envelope in free fall onto the star of mass \Mstar\ at a rate that is 1--3 times the disk accretion rate \Macc. 
%For each of our 4 source models, we have computed \nff(1000 AU), the H$_2$ density at 1000 AU of an envelope in free-fall at rate \Macc\ towards the star of mass \Mstar. The values of \nff(1000 AU) are given in Column 4 of Table~\ref{tab:models}. They scale as \Macc \Mstar$^{-0.5}$ ie. in the same way as the density in the (self-similar) disk wind. 

First of all, Figure~\ref{fig:TdV-nH} shows that the MHD disk wind model adopted here predicts a good correlation  with a slope close to 1 between \TdV(200 pc) and the fiducial envelope density \nff(1000 AU). The latter acts as a proxy for how the disk wind density changes from one model to the next (both quantities scaling as \Macc \Mstar$^{-0.5}$).  Hence this means that {the modeled disk wind luminosity in the 557 GHz line is roughly proportional to the wind density (i.e. mass) inside the HIFI beam over a range of a factor of 45}. 

Second, Figure~\ref{fig:TdV-nH} shows that the model predictions can reproduce very well the observed correlation and the dispersion among datapoints if \nenv(1000 AU) $\simeq 1-3 \times$ \nff(1000 AU); in other words, if the envelope free-fall rate measured at 1000 AU is $\simeq 1-3$ times the disk accretion rate in the MHD wind launching region inside 3--25 AU. Since the disk mass must increase early on, and the final mass of stars appears to be on average one-third of the initial core mass \citep{CMF-IMF}, such a relation between infall and accretion rates does not seem implausible in low-mass Class 0/1 sources. 
%{  check if class1 = 1 and class0 = 3}

%in fact simply a correlation between the wind luminosity at 557 GHz and the wind mass. 

\begin{figure} 
        \centering
        \includegraphics[angle=0, width=\linewidth]{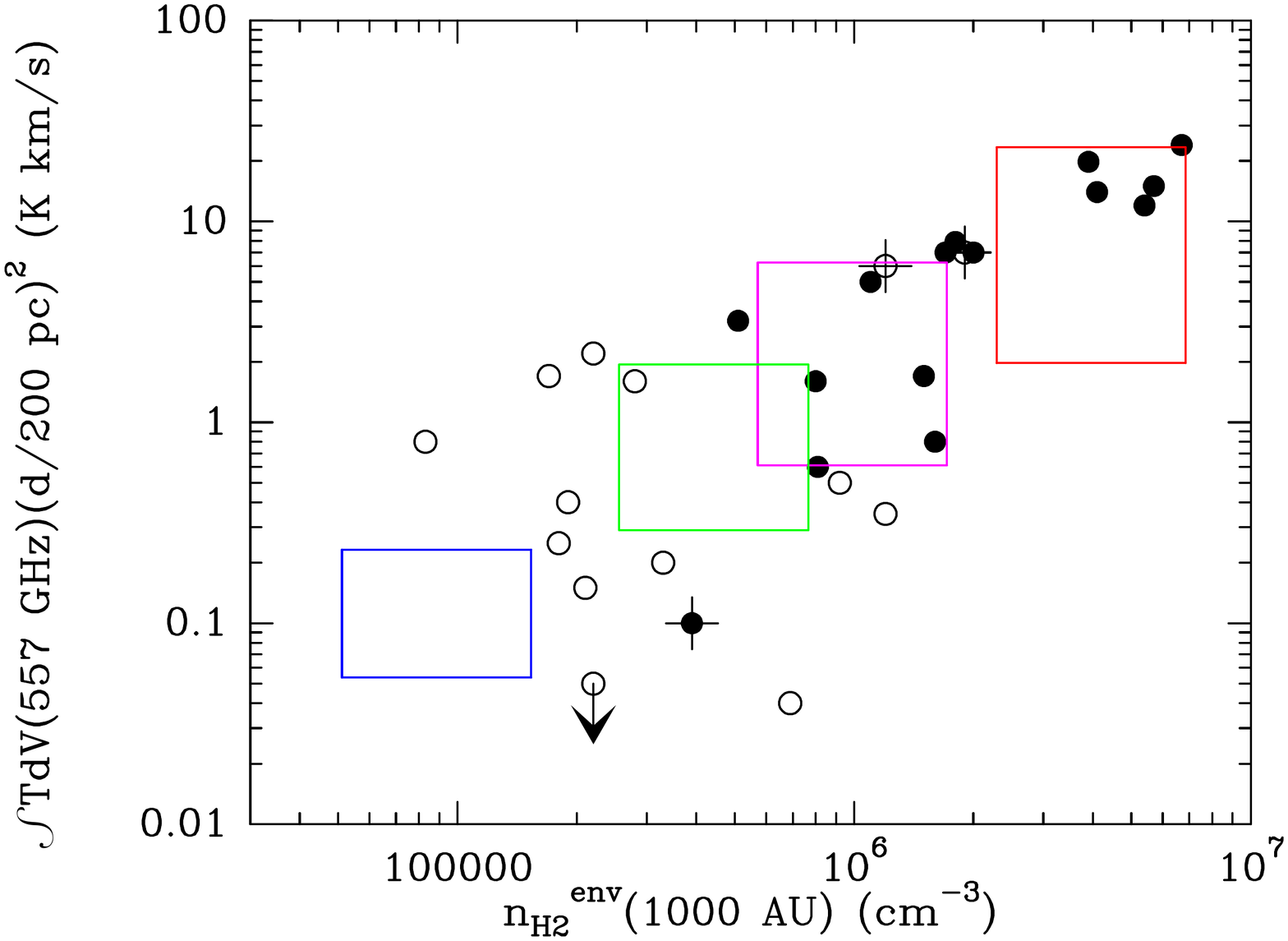}
        \caption{Correlation between \TdV(200 pc), the integrated intensity in the 557 GHz line scaled to a distance of 200 pc, and \nenv(1000 AU), the H$_2$ envelope density at 1000 AU. Data points are from \citet{Kristensen2012557GHz} and labeled according to their $T_{\rm bol}$ classification (filled circles = Class 0, open circles = Class 1).  The borderline Class 0/1 sources according to our 557 GHz line modeling (see text) are indicated with a vertical cross. Predictions for the four models in Table~\ref{tab:models} are represented as rectangles: $red$ = X0, $pink$ = S0, $green$ = X1, $blue$ = S1. The vertical bars illustrate the \TdV\ range for \romax = 3.2 -- 25 AU, and $i= 30\degr - 80\degr$. Horizontal bars illustrate the range of \nenv(1000 AU) assuming an envelope in free fall at a rate of 1--3 times the disk accretion rate in each model. 
        %The black solid line connects the results for \romax = 25 AU and $i = 60\degr$.   
        %{  Their best linear fit to the data points is shown with a dashed line}. 
        }
        \label{fig:TdV-nH}
\end{figure}

\section{Discussion}\label{section_discussion}
%--------------------------------------------------------------------------------------------------%

Our modeling results have shown that a model of dusty magneto-centrifugal disk wind is able to reproduce impressively well the range of shapes and intensities of 
the broad component of various H$_2$O lines in low-mass Class 0 and Class 1 sources, as observed with HIFI/{\em Herschel} by \citet{Kristensen2012557GHz}. It can also reproduce well the close correlation between 557GHz line intensity and envelope density discovered by these authors, provided the infall rate at 1000 AU is $\simeq 1-3$ times the disk accretion rate. As a consistency check, we now examine whether such an interpretation is compatible, at least in an average sense, with independent estimates of disk properties and system inclinations in low-mass protostars.

\subsection{Disk radii in low-mass protostars}
\label{sec:diskobs}

If the broad component of H$_2$O in low-mass protostars does arise predominantly in a warm molecular MHD disk wind, the present exploratory models suggest that low-mass Class 0 and Class 1 protostars should all possess a Keplerian accretion disk, with sufficient magnetic flux to launch magneto-centrifugal ejection out to $\simeq 3-25$ AU. Pseudo-disks where motions are essentially radial rather than toroidal (due to magnetic braking on larger scales) would not be expected to generate enough magnetic twist to drive a powerful magneto-centrifugal outflow until the infalling matter approaches the centrifugal barrier. 
% and (2) accretion rates $\simeq$ 30\%--100\% of the envelope infall rate. 
Here we briefly discuss whether this requirement seems consistent with the known Keplerian disk sizes in low-mass Class 0/1 protostars. 

Direct observational evidence for Keplerian disks around embedded protostars has long been challenging to obtain because of heavy confusion with the dense infalling and rotating large-scale envelope. However, the continued improvement in angular resolution and sensitivity of radio interferometers has revealed several good candidates among both Class 1 and Class 0 sources (see the compilation in Table 7 of \citet{Harsono2014} and the more recent ALMA results of \citep{Sakai2014,L1527-Ohashi2014,Lee2014-HH212,Codella2014-HH212}). 
%{  (CHECK latest papers): 
%IRAS4A2: 220 $\AU$ \citep{Choi-NH3-IRAS4A-2010}, VLA1623 \citep{Murillo2013}. HH212.}

The inferred outer radii for these Keplerian disks are generally 50--200 AU, comfortably larger than the maximum ejection radius \romax\ $\simeq 3-25$ AU needed to reproduce the entire broad component of the H$_2$O 557GHz line with the present MHD disk wind model (see Table\ref{tab:fit}). The most stringent constraint for our model is provided by the close binary system of L1551-IRS5 where individual circumstellar disks appear tidally truncated at an outer radius of only 8 AU \citep{Lim-Takakuwa06}. Our best grid model for L1551 has \romax\ = 3.2 AU, which is consistent with this limit. 

%{  show histogram of \romax\  for class 0 and class 1; do we see any evolutionary trend ? Mention \Mstar\ values inferred from kepler laws? }. 

%Note that the stellar masses inferred by fitting keplerian rotation curves to the interferometric disk observations are in the same range as the 
%\Mstar\ values adopted for our models. {  give specific list here}. 
%Higher values of \Mstar\ have been inferred in some sources from modeling of inverse P Cygni signatures in H$_2$O (Joe's paper), which involved a number of assumptions and several free parameters. Keplerian rotation measurements would be desirable to confirm these values.  

\subsection{Disk accretion rates in low-mass protostars}
\label{sec:maccobs}

Disk accretion rates in Class 0/1 protostars remain highly uncertain. {  In such embedded sources,} empirical proxies for \Lacc\ calibrated on Class 2 disks (FUV excess from the accretion shock, Br$\gamma$ luminosity, etc.) suffer extremely high circumstellar extinction {  and scattering} that cannot be accurately corrected for. 
%%depends on the (unknown) viewing angle, 3D geometry, and dust optical properties. The absorbed photons are re-emitted at longer wavelengths, hence one might hope that the apparent bolometric luminosity \Lbol\ gives some estimate of the total accretion luminosity \Lacc.  (obtained by multiplying by $4\pi$ the bolometric flux seen from earth). The presence of a disk and of scattered light escaping from outflow cavities means that even the bolometric flux (integrated over all wavelengths) will vary with inclination. and (Robitaille). and makes the bolometric flux vary with inclination. 
Since the energy of absorbed FUV and near-IR photons is eventually re-emitted at longer wavelengths, the bolometric luminosity \Lbol (integrated over all wavelengths) may offer a better proxy for the accretion luminosity \Lacc\ in low-mass protostars. However, it still involves large uncertainties due to the unknown fraction $\varepsilon_{\rm rad}$ of accretion power that is actually radiated away instead of being used to drive ejection(s) or heat the star \citep{Adams-Shu}, and to the ill-known stellar photospheric contribution \Lstar\ in heavily accreting stars. An additional complication is that the dust radiation field is highly anisotropic due to disk inclination, occultation, and photon scattering in bipolar outflow cavities. Therefore, the observed spectral energy distribution (SED) will strongly depend on view angle \citep[see e.g.][]{Robitaille07a}, and the apparent \Lbol\ obtained by multiplying by $4\pi$ the bolometric flux as seen from Earth (which is the only observed quantity) will differ from the true \Lbol\ actually radiated by the source over all directions. For example, the apparent \Lbol\ for L1551-IRS5 is 22$\Lsun$, but its SED can be well fitted by models with a {true} \Lbol\ ranging from 7$\Lsun$ to 72$\Lsun$ \citep[see Table~6 of][]{Robitaille07b}, i.e. up to a factor of 3 smaller or larger than the apparent \Lbol. The full uncertainty range in the {true} \Lbol\ can reach a factor of 50-100 in some sources \citep[see Table~6 of][]{Robitaille07b}. 

Computing angle-dependent SEDs for all our grid models would require complex 2D dust radiative transfer calculations and infalling envelope models that are well beyond the scope of the present work. For the sake of comparing our model-predicted disk accretion rates with observational constraints, we will assume that inclination effects affecting the apparent \Lbol\ will tend to average out over a large sample so that {  the angle-averaged \Lbol\ for our models will be close to the true \Lbol. For the latter,} we consider two simplified illustrative cases: (1) \Lbol = 2$L_\odot$+\Lacc, corresponding to a photosphere of 2$L_\odot$ and 100\% of the accretion luminosity eventually getting radiated ($\varepsilon_{\rm rad} = 1$), and  (2) \Lbol\ = \Lacc/2, corresponding to the situation where \Lstar $\ll$ \Lacc\ (true protostar) and the MHD disk wind extracts all the mechanical energy $\simeq 0.5$\Lacc\ released by accretion through the Keplerian disk, while the rest is radiated by the hot spot ($\varepsilon_{\rm rad} = 0.5$). 

In Figure~\ref{fig:TdV-Lbol}, we plot as rectangles the predicted average location of our four source models in the  \TdV(200 pc) -- \Lbol\ plane, for these two cases. It can be seen that they agree very well with the average \Lbol\ values of the datapoints of \citet{Kristensen2012557GHz}. Furthermore, it is striking that the { systematic  vertical shift between Class 0 and Class 1 sources} observed in this graph  is also  reproduced well. This shift occurs in our disk-wind models because for a given \Lacc, Class 0 sources have a {  5 times} smaller \Mstar\ and larger \Macc\ than Class 1 sources. Hence their disk wind is much denser ($\propto$ \Macc \Mstar$^{-0.5}$) than in Class 1 sources of the same \Lbol, and thus much brighter in the 557GHz line (cf. Fig. \ref{fig:TdV-nH}).

\begin{figure} 
        \centering
        \includegraphics[angle=0, width=\linewidth]{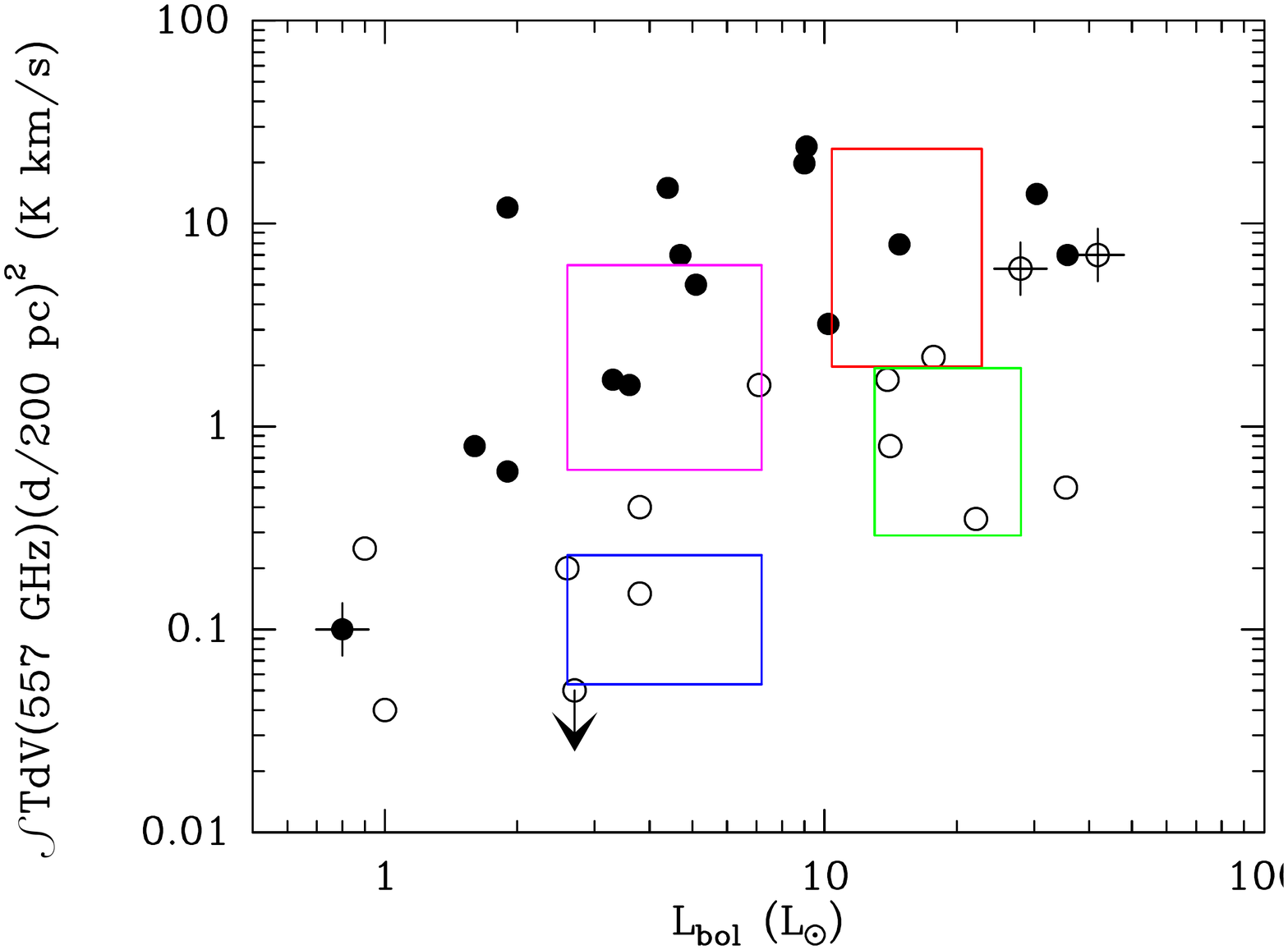}
        \caption{Correlation between \TdV(200 pc), the integrated intensity in the 557 GHz line scaled to a distance of 200 pc, and \Lbol, the apparent bolometric luminosity of the source. Data points from \citet{Kristensen2012557GHz} are  marked according to their $T_{\rm bol}$ classification (filled = Class 0, open = Class 1). The borderline Class 0/Class 1 sources according to our model fits (see text) are indicated by a vertical cross. Colored rectangles illustrate the predicted loci of the four models in Table~\ref{tab:models} for \romax = 3.2 -- 25 AU, $i= 30\degr - 80\degr$, and true \Lbol\ ranging from 0.5\Lacc\ to \Lacc+2$L_\odot$ (see text):  $red$ = X0, $pink$ = S0, $green$ = X1, $blue$ = S1. }
        \label{fig:TdV-Lbol}
\end{figure}

{  We note that the data points span a broader range of apparent \Lbol\ than our model predictions. This is to be expected for several reasons}. First, the angle-dependence of the dust radiation field (due to disk inclination and scattering in outflow cavities) will make the apparent \Lbol\ differ by up to a factor of  3--10 (see discussion above) from the true \Lbol\ that we have used to place our models in Fig.~\ref{fig:TdV-Lbol}. Second, because  \Lacc\ is proportional to \Macc \Mstar, it is much more sensitive to a joint increase in both parameters than the H$_2$O 557GHz luminosity, which scales as  $\simeq$ \Macc \Mstar$^{-0.5}$ (like the envelope density, see Section~\ref{sec:TdV-nH}). For example, an increase in both \Macc\ and \Mstar\ by a factor of 2 will increase \Lacc\ by a factor of 4, but will increase \TdV(200pc) by only $\sqrt{2}$, i.e. 40\%. Therefore, if the observed targets span a slightly broader range of  \Macc\
and \Mstar\ than we have considered in our limited model grid (which is quite
likely), a broader range of \Lbol\ than our model predictions could also easily arise { without  affecting the predicted 557GHz line luminosities} very much. Finally, the accretion shock energy may not be entirely radiated away but used to drive a stellar wind that spins down the protostar \citep{Matt08}, in which case $\varepsilon_{\rm rad} \ll 0.5$ and the true \Lbol\ for our models could extend to smaller values than the range plotted in Fig. \ref{fig:TdV-nH}. We conclude that the MHD disk wind models that reproduce the H$_2$O broad line component in low-mass Class 0/1 protostars involve disk accretion rates that appear consistent on average with observational constraints, given the current observational and theoretical uncertainties in relating the apparent \Lbol\ with the true \Lacc.

\subsection{Inclination angles}
\label{sec:inclobs}
Although our very
sparse parameter grid means that our line profile modeling is only illustrative, it is instructive to look at the statistical distribution of viewing angles suggested by the present modeling exercise. The behavior is quite different between Class 0 and Class 1 sources. Among the 16 sources best matched by our Class 0 grid models (including HH46-IRS and SVS13), the three inclination bins in our grid are equally represented (5--6 sources in each). In contrast, among the 13 sources best matched by our Class 1 grid models, only one is reproduced with $i = 30\degr$ (TMC1A) and two with $i = 60\degr$ (Oph-IRS63 and Ced 110), while the remaining ten require $i = 80\degr$. This reflects the trend for narrower H$_2$O line wings in Class 1 sources already noted by \citet{Kristensen2012557GHz,Mottram2014}.  Our modeling confirms that this difference in line width between the two Classes is real and is not due to signal-to-noise issues; either the Class 1 sample is intrinsically poor in objects seen closer to pole-on (e.g. because these sources would look more like Class 2), or our disk-wind modeling in Class 1 sources predicts excessive H$_2$O line widths, thus requiring artificially high inclinations to fit the observed line profiles. 

To try and distinguish between these two possibilities, we have compared the inclinations of our best matching grid models with independent inclination estimates from jet proper motions (L1448-MM, IRAS4B, HH46-IRS, SVS13), radial velocity (IRAS15398), and bowshock modeling (L1551) or from modeling of the CO outflow kinematics and scattered light images (B335, L1157, L1527, GSS30, L1489, RNO91). These inclination values and the corresponding bibliographic references are listed in the last two columns of Table~\ref{tab:fit}.  Inclination estimates from the mere visual appearance of CO outflow lobes in single-dish maps were considered too uncertain for such a comparison because of  the limited angular resolution\footnote{For example, \citet{Yildiz2012NGC1333} suggest $ i=$ 15\degr--30\degr\ for IRAS4B, whereas proper motions $\simeq 10-40$ \kms\ and radial velocities $\simeq \pm 8$\kms\ of H$_2$O masers \citep{Desmurs09} indicate $i=$ 50\degr--80\degr\ on small-scales. \citet{vanKempen2009} suggest that the CO outflow in IRAS15398 is viewed 15\degr\ from pole-on, whereas the low radial velocity  $\simeq -46$ \kms\ of the associated Herbig-Haro object \citep{Heyer-Graham89} compared to typical jet speeds $\simeq 150$ \kms\ favors a high inclination from the line of sight, $i \ge 70\degr$.  \label{foot:iras15398}}. Inclination estimates derived solely from fits to the source SED were also deemed too uncertain given the many free parameters involved \citep[see e.g. Table 1 of][]{Robitaille07b} and the uncertainties in dust properties. 

Among the 16 sources modeled by Class 0 profiles, 7 have independent inclination estimates and except for L1448-MM (which is not well fitted by any of our grid models) they are all consistent with our modeled inclination to within 10\degr, which is reassuring. Among the 13 sources modeled by Class 1 profiles, 
%two (TMR1 and TMC1) have only loosely constrained inclinations of 20\degr--80\degr\ from mapping of their rotating disks  \citep{Harsono} and 
only 4 have direct inclination estimates: 35\degr--55\degr in L1551, $70\degr \pm 4\degr$ in RNO91,  $\simeq 65\degr$ in GSS30, and $\ge 57\degr$ in L1489\footnote{\citet{Brinch2007b} have argued that  the high mid-infrared flux in L1489 requires a much smaller inclination for the disk (40\degr) than for the outflow cavities. However, their SED models do not include internal disk heating by accretion. Furthermore, the dust scattering efficiency at mid-infrared wavelengths would be much increased if large micron-sized grains are present, as suggested by the recent discovery of ``coreshine'' in dense cores \citep{coreshine}. Both effects would increase the predicted mid-IR disk flux and allow for a higher disk inclination}. The 557GHz line profiles of these four sources are all best matched by our Class 1 grid models with $i = 80\degr$ (see Fig.~\ref{fig:BestFit}). Hence, our Class 1 models do predict excessive inclination in two out of the four cases. Although the statistics are poor, they do suggest that our models for Class 1 sources seem to predict line wings that are slightly too broad. Several situations could lead to narrower line wings in Class 1 MHD disk winds than in our present models: (i) the initial H$_2$O abundance on inner streamlines is smaller than assumed, (ii) the MHD ejection starts well beyond \Rsub\  so that only slower streamlines emit in H$_2$O, or (iii) the wind magnetic lever arm is smaller than in Class 0 sources, hence the wind is accelerated to lower speeds. 
%It is of course also possible that the 557 GHz line profiles in Class 1 sources become dominated by a different component of intrinsically lower velocity, eg. shocked and irradiated swept-up gas in the outflow. 
Testing each of these options requires dedicated modeling that lies beyond the scope of the present exploratory work, and will be the subject of a forthcoming article. 

%\subsection{Model limitations}

%TODO : repomper quelques § de la thèse ici

%\NEW{The class 1 model with $\Macc = 10^{-6} \, \UMacc$ has profile
%intensity lower than 15~$\mathrm{mK}$, hence this model could not be
%present in Figure~\ref{FIGURETDVversusL}. While the class 1 model
%with $\Macc = 5 \times 10^{-6} \, \UMacc$ could have model points in this
%figure for $\romax > 12.8 \, \AU$ only, but degenerated with
%predicted point from class 0 model with same accretion rate and
%$3.2 \, \AU < \romax < 6.4 \, \AU$.}

%--------------------------------------------------------------------------------------------------%
\section{Conclusions}\label{sec:concl}
%--------------------------------------------------------------------------------------------------%

The calculations presented in this paper show that the MHD disk wind model invoked to explain jet rotation in T Tauri stars, if 
 launched from dusty accretion disks around low-mass Class 0 and Class 1 protostars, would be sufficiently warm, dense, and water rich to produce H$_2$O emission detectable by HIFI on board the {\it Herschel} satellite. 
%We find that pumping by the dust submm background can have a significant effect on the H$_2$O excitation near the wind base, resulting in net gas heating by collisional de-excitation. 
We find that the H$_2$O line profiles predicted by this model reproduce the observational data remarkably well  in spite of  the sparse grid of parameters (\Macc\, \Mstar\,\romax, inclination) that we have considered in our model grid. In particular, they can reproduce 
%While we consider only one particular MHD solution, and our results are only illustrative and not meant to fit each source in detail, we find that a reasonable range of parameters (\Macc\, \Mstar\,\romax, inclination) can reproduce very well at the same time several characteristics of the H$_2$O line emission in low-mass Class 0 and Class 1 protostars as observed by \citet{Kristensen2012557GHz}:
\begin{itemize} 
\item the shape, velocity extent, and intensity of the broad component of the fundamental o-H$_2$O line at 557GHz;
\item the profiles and relative intensities of this broad component in more excited H$_2$O lines;
\item the good correlation with a slope $\simeq 1$ between the H$_2$O 557GHz luminosity and the envelope density at 1000 AU;
\item the systematically higher water luminosity of Class 0 sources compared to  Class 1 sources of the same bolometric luminosity.  
\end{itemize}
These results indicate that MHD disk winds are an interesting option to consider for the origin of the broad H$_2$O line component discovered by \citet{Kristensen2012557GHz} towards low-mass protostars with HIFI/Herschel. If this interpretation is correct, an important implication would be that all low-luminosity Class 0 protostars  
 possess Keplerian accretion disks that are sufficiently magnetized out to 3--25 AU to launch MHD disk winds that extract a large fraction of the angular momentum necessary for accretion
 and have accretion rates on the  order of 30\%--100\% of their envelope infall rate at 1000 AU.

%We have verified that these requirements appear compatible with current observational constraints on disks in low-mass protostars. 
In the case of Class 1 sources, their systematically narrower line wings in the 557 GHz line, first noted by \citet{Kristensen2012557GHz}, would imply that they are mainly viewed close to edge-on ($i \simeq 80\degr$) or, more likely, that their MHD disk winds (if present) are less water rich and/or slower in their inner regions than the specific MHD solution used here. 
%To distinguish between these options would require accurate independent inclination estimates in more Class 1 sources of this sample (eg. from HST imaging of scattered light cavities, jet proper motions, or disk imaging). 

Of course, the success of the MHD disk wind model explored here in fitting the broad component of HIFI/{\it Herschel} H$_2$O line profiles does not mean that this model is necessarily correct {  or unique. Other possible origins for the broad component have been proposed that are not ruled out at this stage}, such as shocked material or mixing-layers along the outflow cavity walls \citep[see e.g.][]{Mottram2014}. Further tests of the disk wind hypothesis need to be undertaken in a range of molecular tracers {  (including e.g. CO $J=16-15$ HIFI line profiles, Kristensen et al., in prep.)} and with higher angular resolution. The revolutionary mapping sensitivity and angular resolution of the ALMA instrument are particularly well suited to searching for the slow and warm wide-angle component predicted by MHD disk wind models. Synthetic predictions for {  CO and} early-science ALMA will be presented in a forthcoming paper and compared with available data. 

%--------------------------------------------------------------------------------------------------%
\begin{acknowledgements}
{  We are grateful to Simon Bruderer, Moshe Elitzur, David Flower, Lars Kristensen, Joe Mottram, and Laurent Pagani for useful discussion and advice at various stages of this work, and to the anonymous referee for constructive comments that helped to improve the paper. This work received financial support from the INSU/CNRS Programme National de Physique et Chimie du Milieu Interstellaire (PCMI), and made use of NASA's Astrophysics Data System.}
\end{acknowledgements}
%--------------------------------------------------------------------------------------------------%

\bibliographystyle{aa}
\bibliography{BIBLIOGRAPHIE}

%--------------------------------------------------------------------------------------------------%
%--------------------------------------------------------------------------------------------------%
\appendix
%--------------------------------------------------------------------------------------------------%
%--------------------------------------------------------------------------------------------------%

%--------------------------------------------------------------------------------------------------%
%\section{Model improvements}\label{app_model}
%--------------------------------------------------------------------------------------------------%

\section{$\HH$ and CO self-shielding} 
\label{app:shielding}

One improvement in our model calculations since \citet{Panoglou2012} consists in the implementation of an accurate estimation of the shielding column of $\HH$ molecules along the line of sight to the star, $\N_{\HH}$.
We now calculate the streamlines evolution in increasing order of
launching radius $r_0$, starting from the dust sublimation radius \Rsub, where we assume $\N_{\HH} = 0$.
The increment in $\HH$ column density between two consecutive streamlines $i-1$ and $i$
at each polar angle $\theta$ is approximated by~
%\begin{align}
\begin{equation}
\Delta \N_{\HH} (i,\theta) \simeq  x_{\HH}(i-1,\theta) \times \Delta \N_{\mathrm{H}} (i,\theta), \notag \\ 
%                & =  2 x_{\HH}(i-1,\theta) \, \nH(i,\theta) \, R(i,\theta) \times \Big( \left( \frac{r_0(i)}{r_0(i-1)} \right)^{0.5} - 1 \Big)
%\end{align}
\end{equation}
where $x_{\HH}(i-1,\theta)$ is the fractional abundance of $\HH$ at angle $\theta$ on
the $(i-1)^{\mathrm{th}}$ streamline, %$\nH(i,\theta)$ and $R(i,\theta)$ are the H nucleus density and the spherical radius
%at angle $\theta$ on the $i^{\mathrm{th}}$ streamline, 
and $\Delta \N_{\mathrm{H}} (i,\theta)$ is the increment in column density of H nuclei between streamlines $i-1$ and $i$ along angle $\theta$. We use the $\HH$ abundance on streamline $i-1$ rather than some average value
between the two consecutive streamlines because we found that an average leads to overestimating the shielding when the gradient in $x_{\HH}$ is steep, which will be the case near the H-H$_2$ transition. %$\Delta \N_{\mathrm{H}} (i,\theta)$ has a simple analytical expression as \nH(R,\theta) \propto R^{-1.5}$ for fixed $\theta$.

As we refine the spacing between successive streamlines, the variation of $x_{\HH}(i,\theta)$ as a function of $r_0$ converges to a limiting curve, as illustrated in Figure~\ref{fig:shielding} in the case of our reference S0 model. %at the recollimation point ($z/r = 20$). 
The final adopted discretization in $r_0$ results from a trade-off between accuracy in $x_{\HH}$ and CPU time. 
The same method is applied to calculate the shielding column of CO molecules along the line of sight to the star, $\N_{\rm CO}$. 

   \begin{figure}
   \centering
   \begin{tabular}{c}
   \includegraphics[width=\linewidth]{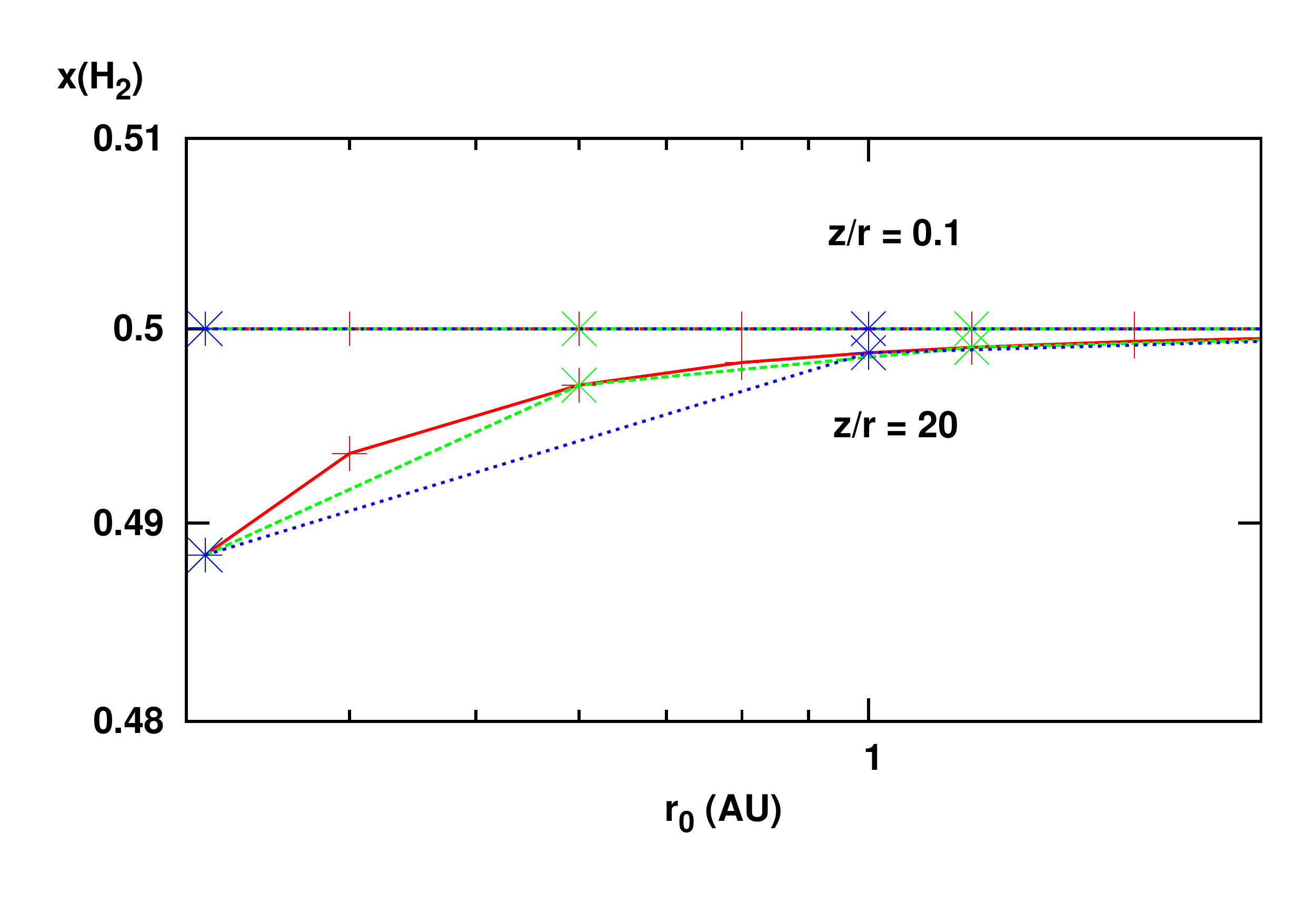} \\
   %\includegraphics[width=\linewidth]{H2_DISCRETISATION2} \\
   %gs -q -dSAFER -dNOPAUSE -dEPSFitPage -dBATCH -sOutputFile=Shielding.pdf -dDEVICEWIDTHPOINTS=500 -dDEVICEHEIGHTPOINTS=720 -sDEVICE=pdfwrite -c '.setpdfwrite <</PageOffset [-50 -50]>> setpagedevice' -f Shielding.ps
   \end{tabular}
   \caption{Effect of a finer sampling in streamline radii $r_0$ on the calculated fractional H$_2$ abundance $x_{\HH} = n(\HH)$ / \nH\ at the recollimation point ($z/r = 20$)  of each streamline for our reference S0 model. We note the convergence to a limiting curve as the sampling is increased and the calculation of the self-shielding H$_2$ column becomes more accurate. }
   %   And for \textbf{class 2} ($bottom$) ($\Mstar = 0.5$ \Msun\ and $\Macc = 10^{-7}$ $\UMacc$)
%   where convergence to the transition to molecular hydrogen is more sensible to the discretization in $r_0$.
%   The curve is more flattened at lower angle $z/r$. }
   \label{fig:shielding}
   \end{figure}

%The limiting curve at small $r_0$
%corresponds to equilibrium between
%photodissociation and reformation on grains.
%At values of $r_0$ very close to $\Rsub$, the finer discretization
%tends to amplify photodissociation of $\HH$,
%but overall it tends to increase shielding and the transition to molecular
%streamlines become sharper and occurs at $r_0$ values closer to $\Rsub$.
%Furthermore, the higher the density,
%the greater the shielding is.
%Hence for class 1 and 0 protostars the limiting curve is flatter than
%than for class 2, and less dependent on $r_0$ sampling.

%---------------------------------------------------------------------%
\section{$\water$ rotational excitation and gas heating/cooling} 
\label{app:lines}

The out-of-equilibrium evolution of H$_2$O rotational level populations
along each streamline is calculated under the large velocity gradient (LVG) approximation
\citep{Sobolev57} using the method developed for planar C-shocks by \citet{Flower2010H2O}.
We consider the first 90 rotational levels of $\water$ (45 ortho and 45 para),
with energies up to $2\,000$ K taken from \citet{Tenysson2001}, and their coupling
by 320 radiative transitions with $A_{ij}$ from \citet{Barber2006}. 
The rate coefficients adopted for collisions with He and H are the same as in \citet{Flower2010H2O}. 
%with He from \citet{Green93}, 
%while those with H are the same as for excitation by ortho-H$_2$. 
For collisions with H$_2$ we use the more recent rate coefficients computed by \citet{Dubernet2009,Daniel2010}. 
When the kinetic temperature exceeds the range where the coefficients are calculated, 
we use the last tabulated value. The rate of gas cooling/heating by H$_2$O is calculated self-consistently with the 
local H$_2$O level populations, as the net rate of energy lost/gained by the gas through $\water$ collisional excitation or de-excitation. 
This differs from Paper I, where cooling by H$_2$O was evaluated from the tabulated LVG cooling functions of \citet{NK93}.

%{\em Collision rates:} 
%For the same reasons as in \citet{Flower2012outflow},
%we adopt for ortho-$\HH$ collisions the rate coefficients with $\He$, scaled by the reduced mass.
%{  Do we really do this ?? I thought that Dubernet and Daniel also provided coeffs with o-H2 ?}.

%---------------------------------------------------------------------%
%\section{Escape probability calculation} \label{model_vgrad}

Unlike the shock models of \citet{Flower2010H2O}, the disk wind is not plane-parallel. 
Hence the variation of photon escape probability with angle must be evaluated numerically.
The projected velocity gradient $\frac{\partial v_s}{\partial s}(\theta_1,\phi_1)$
in each direction $(\theta_1, \phi_1)$ is calculated numerically by interpolating the velocity field in the MHD solution.
The Sobolev optical depth of the transition $u \rightarrow l$ in this direction is given by~\citep[see e.g.][]{Surdej1977}
\begin{equation}
\tau_{ul}(\theta_1,\phi_1) = \frac{n_u A_{ul} c^3 }{8 \pi \nu_{ul}^3\vert{\frac{\partial v_s}{\partial s}} (\theta_1,\phi_1)\vert} 
\left(\frac{g_u n_l}{g_l n_u} - 1\right)
,\end{equation}
where $g$ and $n$ denote the level statistical weight and number population,
and $\nu_{ul}$ and $A_{ul}$ are the transition frequency and spontaneous radiative decay rate.
The corresponding photon escape probability is $\beta(\theta_1,\phi_1) = (1-e^{-\tau_{ul}}) / \tau_{ul}$. 
A few lines occasionally show a maser effect with $\tau_{ul} < 0$, so that $\beta_{ul}$ becomes greater than 1. Since the LVG treatment does not include maser saturation, we impose in this case an arbitrary upper limit of $\beta_{\rm max} = 5$. However, this occurs only in very few directions and positions, and does not affect the emergent spectra. 

The mean escape probability $<\beta>$ averaged over all solid angles is obtained from Gaussian integration of $\beta(\theta_1,\phi_1)$ over 12 values in $\theta_1$ and 12 values in $\phi_1$. The mean radiation field at the line frequency is then calculated with the usual LVG formula 
\citep[e.g.][]{Surdej1977} 
\begin{equation}
\bar{J}_{ul} = (1-<\beta>) B_{\nu_{ul}}(T_{\rm ex}) + <\beta> \bar{J}^{\rm bcg}_{\nu_{ul}}, 
\end{equation}
where $T_{\rm ex}$ is the excitation temperature of the transition, 
and $\bar{J}^{\rm bcg}_{\nu_{ul}}$ is the background radiation field at the line frequency. The latter
is taken equal to $\bar{J}^{\rm dust}_{\nu_{ul}}$, the diffuse radiation field in the dusty disk wind as computed by the DUSTY code (see Appendix~\ref{app:model_DUSTY}). The value of $\bar{J}_{ul}$ is then used to calculate the rates of change of the level populations by stimulated emission and absorption, and evolve them to the next point in $z$ along the streamline. The calculated escape probability in each direction $(\theta_1,\phi_1)$ is saved and used to compute the contribution of each gas cell to the final emergent spectra. 
\section{Dust temperature and diffuse background} \label{app:model_DUSTY}

In Paper~I, the grain temperature was computed from radiation equilibrium against the direct stellar and hot-spot radiation, attenuated by intervening dusty wind streamlines. This method is valid as long as the dust is optically thin to its own infrared radiation, an assumption that is not fully adequate in the dense winds from Class 1 and Class 0 sources, where backwarming becomes significant. We now include this effect in an approximate way, using the 1D radiative transfer code DUSTY \citep{DUSTY1999}.  We adopt the same standard MRN grain composition, optical properties and size distribution as in Paper I. 
We approximate the density of H nuclei in the wind as a 1D power-law distribution 
\nH$(r,\theta) \simeq 1/20 \times$ \nff($R$),
%1/55 \times n_\mathrm{ff}(R)$
which provides a good approximation over the intermediate streamline region between the base and the recollimation point. %(see Figure~\ref{fig:DUSTY_Tdust}).
Here $R$ is the spherical radius and \nff\ is the fiducial H$_2$ density in a spherical envelope 
in free fall at rate \Macc\ onto a point source of mass \Mstar
\begin{eqnarray}
n_\mathrm{H_2}^{\rm acc}(R) & \equiv &\frac{1}{2} \frac{\mMacc R^{-1.5}}{4 \pi \mu m_\mathrm{p} \sqrt{2 G \mMstar}}  \\
%\simeq 1.6 \times 10^6 {\rm cm}^{-3}
% NB this factor must be divided by 1.4 for Helium and by 2 to give H2 density
&\simeq& 6 \times 10^5 {\rm cm}^{-3}
\left(\frac{\mMacc}{5\times 10^{-6} \UMacc} \right) \left(\frac{\mMstar} {0.1 M_\odot} \right)^{-0.5} \\
& & \times \left(\frac{R}{1000\, \AU} \right)^{-1.5} 
\label{eq:nff}
,\end{eqnarray}
where $\mu = 1.4$ accounts for the mass in the form of helium.

We take the inner radius of the 1D dust shell to be at the sublimation radius, denoted \Rsub, where \Tdust = 1500~$\K$. 
%The 1D radial opacity of the "shell" at $\lambda$ = 5500\AA integrated from $\Rsub$ to $R_{\rm max} \gg R_{\rm sub}$ is then~:
%\begin{align}
%\tau_0 & = \frac{1}{55} 2 n_\mathrm{ff}(\Rsub) \Rsub  \sigma_{\rm H}^d(5500 \AA) %= \frac{1}{55} \times \frac{\mMacc \sigma_d(5500 \AAr) }{2 \pi m_P \Big(2 G \mMstar \, \Rsub \Big)^{0.5}}, \notag \\
%       & = 13.4 \left( \frac{\mMacc}{5 \times 10^{-6} \, \UMacc} \right) \left( \frac{\mMstar}{0.1 \, \mMsun} \frac{\Rsub}{0.31 \, \AU}\right)^{-0.5}
%\end{align}
%where $\sigma_{\rm H}(5500\, \AA) = 2.72 \times 10^{-22} \, {\rm cm}^{2}$ is the dust absorption cross section per H nucleus at 5500\AA\ for our dust model (see Paper I).
The dust is heated by a central source made of two blackbodies, a stellar photosphere of 2 $\Lsun$ at 4000~K, and an accretion shock at 10,000 K radiating half of the accretion luminosity~(see Paper~I for a discussion of these assumptions).
% one at 4\,000~$\K$ representing
%the stellar photosphere with $\Rstar = 3 \, \Rsun$, $L_\star = 2.1 \Lsun$,
%and another at 10\,000~$\K$ representing the accretion hot spots with
%luminosity $L_\mathrm{hs} = \frac{1}{2} G \mMstar \mMacc / \Rstar$ = $2.6 L_\odot %\times \left({\mMstar}/{0.1\, \mMsun} \right) 
%\left({\mMacc}/{5 \times 10^{-6} \, \UMacc}\right)$. 
%Because the total dust opacity of the wind depends on \Rsub\ and reciprocally, a few iterations of DUSTY are needed to reach convergence. 
The calculated dust sublimation radius is listed in Table~\ref{tab:models} for our four source models. It is $\simeq 0.3 \AU$ in models S0 and S1 
(50\% larger than in Paper I), and $\simeq 0.6 \AU$ in models X0 and X1.  
%is $\Rsub(\tau_0 = 13) = 0.31 \,\AU$ in the class 0 model,
%close to the 0.2~$\AU$ from Paper I. When $\Macc = 2 \times 10^{-5} \,
%\UMacc$, $\tau_0 = 52$, $L_\mathrm{hs} = 10.5 \, \Lsun$
%$\Rsub$ increases to 0.63~$\AU$.
%The calculated 1D dust temperature profiles for our 4 source models are plotted in Figure~\ref{fig:DUSTY_Tdust}. 
%The temperature initially falls off quite rapidly out to 10 AU and
%then follows a power law $T_{\rm dust} \propto R^{-0.36}$ for $R > 10 \,\AU$.
%The dust temperature differs among the 4 models by a factor 1.5 at each radius, despite the range of a factor 5 in hot spot luminosities and 45 in wind densities, suggesting that $T_{\rm dust}$ should not be too strongly affected by our simplifying approximations. 
\citet{Visser2009} carried out full 2D dust radiative transfer calculations in a rotating collapsing singular isothermal sphere and found that the disk surface reaches a dust temperature of 100 K out to 17--41 AU in their reference and standard models. For comparison, we find \Tdust\ = 100 K at 25--80 AU in our 1D DUSTY calculations for a comparable source luminosity. The use of an approximate 1D dust radiative transfer thus appears sufficiently accurate for our purposes, given the other simplifying hypotheses in our model and the exploratory nature of our work.

%with important consequences for the release of water ice into the gas-phase (see Appendix~\ref{app:initial_abun}), and for the gas kinetic temperature and H$_2$O collisional excitation.
%\begin{figure}
%\centering
%%scp -r -p yvart@aramis:/home/yvart/mhdjet_2/Classe0-Jan12/PLOT/Tdsutclass0.ps ./
%%gs -q -dSAFER -dNOPAUSE -dEPSFitPage -dBATCH -sOutputFile=Tdsutclass0.pdf -dDEVICEWIDTHPOINTS=500 -dDEVICEHEIGHTPOINTS=720 -sDEVICE=pdfwrite -c '.setpdfwrite <</PageOffset [-50 -50]>> setpagedevice' -f Tdsutclass0.ps
%\includegraphics[width=7.5 cm, angle=0]{newfigs/DUSTY_density.pdf}
%\caption{
%       H nuclei density \nH\ along a disk wind streamline as a function of spherical radius $R$, and its 1D fit by 1/20 times the spherical free-fall law \nff($R$) from Equ.~{eq:nff}. 
%%      $Right$~: dust temperature as a function of spherical radius $R$ for our 4 source models, 
%%      as calculated by DUSTY using the 1D density power-law shown in the left panel.
%       {  This figure needs to be updated.}}
%\label{fig:DUSTY_Tdust}
%\end{figure}

The DUSTY code also calculates the local angle-averaged specific intensity $\bar{J}_\lambda^{\rm dust}$ \citep[see Equ. C.1 in][]{DUSTY1996}.
%Figure~\ref{fig:DUSTY_Jnu}  plots $J_\lambda^{\rm dust}$ at various radii for the class 0 reference model. 
It includes the submillimeter diffuse radiation emitted by warm dust grains, as well as the attenuated and geometrically diluted contribution of the star and hot spots at shorter wavelengths. This background radiation field is taken into account in computing the $\water$ rotational level excitation, as discussed in Appendix~\ref{app:lines}.

%$\Rsub = 0.27 \, \AU$ in class 1 model with 
%\Macc = $10^{-6} \, \UMacc$, $\tau_0 = 1.2$ and $L_\mathrm{hs} = 2.1 \, \Lsun$.

%   TAU0 = 13
%   Fe1 = 6.04E+04 #W/m2 NEW
%   ypos01 = 1.0       ; T01 = 1500.0
%   ypos10 = 6.081E+00 ; T10 = 350 #3.485E+02
%   ypos15 = 1.946E+02 ; T15 = 80 #7.557E+01
%   ypos20 = 6.227E+03 ; T20 = 20 #2.270E+01

%\begin{figure}
%\centering
%%cp /home/yvart/Classe0/Tdust_Ceccarelli/Jl_Bl_PAPER.ps ./
%%gs -q -dSAFER -dNOPAUSE -dEPSFitPage -dBATCH -sOutputFile=Jl_Bl_PAPER.pdf -dDEVICEWIDTHPOINTS=500 -dDEVICEHEIGHTPOINTS=720 -sDEVICE=pdfwrite -c '.setpdfwrite <</PageOffset [-50 -50]>> setpagedevice' -f Jl_Bl_PAPER.ps
%\includegraphics[width=7.5 cm, angle=0]{newfigs/Jl_Bl_PAPER}
%\caption{Cross symbols: Angle-averaged radiation field $J_\lambda^{\rm dust}$ at various radii  ($R =$ \Rsub\ in $red$, 19~$\AU$ in $green$, 60~$\AU$ in $blue$ and 2\,000~$\AU$ in $purple$) calculated with the 1D DUSTY code in the standard Class 0 model.
%       Thin solid curves plot blackbody spectra at the local dust temperature \Tdust($R$) for comparison.
%       %The silicates are seen in emission.
%}
%\label{fig:DUSTY_Jnu}
%\end{figure}

%---------------------------------------------------------------------%
\section{Initial conditions at the wind base} \label{app:initial_abun}

%An important difference from Paper~I is Inclusion of photon trapping in the 1D DUSTY calculations yields a higher dust temperature, and values of 100 K sufficient for thermal desorption of water ice  \citep{Fraser2001H2O, Oberg2009H2O} are now reached at $R \leq  25 \, \AU$  (see Fig.~\ref{fig:DUSTY_Tdust}). 

%Before starting the thermo-chemical integration along each streamline, 
In Paper I, the initial temperature and chemical abundances at the base of each streamline were obtained in a simple way by calculating the local steady-state thermochemical equilibrium (reached typically after $3 \times 10^5$ yrs). Although convenient, this procedure did not take into account the previous chemical history of the gas as it was accreted from larger radii up to the current $r_0$. When the disk angular momentum is mainly extracted by MHD wind torques (as in the MHD solution adopted here), the accretion flow is much faster than in a viscous disk and typically transonic with $V_{\rm acc} \simeq 0.7 C_s$, where the sound speed $C_s$ is related to the Keplerian speed through $C_s = (h/r_0) V_\mathrm{Kep}(r_0)$. The disk aspect ratio $h/r_0 \equiv \epsilon$ is constant in our self-similar model, with a value of 0.03.  As a consequence, the characteristic radial drift timescale $t_\mathrm{acc}$ at distance $r_0$ from the star is 
%$t_\mathrm{acc} = R/v_\mathrm{Radial}$ with $v_\mathrm{Radial} \simeq C_s$, hence
%we take~:}
\begin{eqnarray}
t_\mathrm{acc}(r_0)  & \simeq & \frac{r_0}{V_{\rm acc}} \simeq \frac{r_0}{0.7 \varepsilon V_\mathrm{Kep}(r_0)} \notag \\
%& \simeq 43 \times r_0^{3/2} (G \mMstar )^{-0.5} \notag \\
& \simeq & \, 22 \yr \times \left( \frac{r_0}{1 \, \AU}\right)^{1.5} \left(\frac{\mMstar}{0.1 \, \mMsun} \right)^{-0.5},
\label{eq:tacc}
\end{eqnarray}
which is much shorter than the timescale to reach chemical equilibrium.

Computing the full 2D non-equilibrium chemical evolution along the accretion flow is a highly complex problem that represents a full work in its own right \cite{Visser2009,Hincelin2013}. 
%(these issues are not present in the pure hydrodynamical case, eg.\citet{Visser2009}). 
%For the purpose of the present exploratory study, we adopt here a simpler approximate approach :
For the purpose of the present exploratory paper, we adopt the following simpler approach to estimating chemical abundances at the wind slow point.
We first start from typical steady-state dark cloud conditions with the ice phase composition of \cite{Flower2003} and evolve the gas-phase chemical abundances at $r_0 = 50 \, \AU$ for a duration of $3 \times 10^{5} \, \yr$, typical of the viscous accretion time in the outer disk where ionization is too low for MHD ejection to take place. This is sufficient to reach chemical steady state, hence the result does not depend on the precise adopted duration. We then propagate the resulting abundances at the wind slow point at $r_0 = 25$AU, putting back in the gas phase all ice species that are thermally desorbed (\Tdust $\ge 100$ K), and to mimic the effect of accretion we evolve the abundances and gas temperature at the local density $n_H$(25 AU) for a short duration equal to the accretion timescale $t_\mathrm{acc}(r_0)$ given by Eq.~\ref{eq:tacc}. The result is taken as initial conditions for the thermo-chemical integration along the 25 AU wind streamline. We then proceed in the same way at the next smaller $r_0$ = 12.8 AU, using the calculated abundances at the slow point at 25 AU as input and evolving them for a duration $t_\mathrm{acc}(r_0)$ to obtain the initial conditions for the thermo-chemical integration along that wind streamline. We progress recursively in this way from outside in, down to $r_0 =$ \Rsub. 

%all ice species are thermally desorbed \citep{Fraser2001H2O, Oberg2009CO, Oberg2009H2O} into the gas phase. 

Table~\ref{tab:initial_abun} summarizes the calculated initial gas and dust temperatures and $\water$ gas-phase abundance at the disk-wind slow point on a few representative streamlines for the standard Class 0 model, together with the local H nucleus density (imposed by the MHD solution). The main difference from Paper~I is that the higher dust temperature, and ensuing ice thermal evaporation leads to a water gas-phase abundance of about $10^{-4}$ at all launch radii $\le 25 \, \AU$. When the gas temperature exceeds 250~K, additional water forms by endo-energetic reaction of O and OH with H$_2$, but the accretion timescale is too short for this conversion to be complete. Similar initial water gas-phase abundances are found for the other three source models in Table~\ref{tab:models}. 

%Level populations of $\water$ are assumed at LTE.

\noindent

\begin{table}
\caption{Initial conditions at the wind slow point for the standard Class 0 model}
\label{tab:initial_abun}
\centering
\begin{tabular}{ l l l l l l}
\hline\hline
$r_0$  & $t_{\mathrm{acc}}$  & $T_{kin}$  & $T_{dust}$  & \nH\  & $n(\water)/n_{\rm H}$  \\
($\AU$)& ($\yr$)             & ($\K$)     & ($\K$)      & (\cm) & gas phase \\
\hline
%50            & $3 \times 10^{5}$                & 70               & 80              & $2 \times 10^{8}$  & $(1.2 \times 10^{-5} )^a$ \\
25            & $\sim$2700                       & 100              & 100              & $5.6 \times 10^{8}$& $1.1 \times 10^{-4}$ \\
12.8          & $\sim$1000                        & 130              & 140               & $1.1 \times 10^{9}$& $1.1 \times 10^{-4}$ \\
6.4           & $\sim$350                        & 180              & 190              & $4.3 \times 10^{9}$& $1.1 \times 10^{-4}$ \\
3.2           & $\sim$125                        & 250              & 260              & $1.2 \times 10^{10}$& $1.2 \times 10^{-4}$ \\
1             & $\sim$22                         & 500              & 510              & $6.9 \times 10^{10}$& $1.2 \times 10^{-4}$ \\
0.31          & $\sim$3.8                        & 1300             & 1500              & $4 \times 10^{11}$& $1.5 \times 10^{-4}$ \\
\hline
\end{tabular}
%\tablefoot{($a$) For $r_0 = 50 \, \AU$,  $x(\water)_\mathrm{ice} = 10^{-4}$ \citep{Flower2003}.}
\end{table}

\end{document}